\documentclass[twocolumn,showpacs,floatfix,preprintnumbers,amsmath,amssymb,floatfix]{revtex4}

\usepackage{graphicx}
\usepackage{dcolumn}
\usepackage{bm}

\begin{document}

\title{Growth, microstructure, and failure of crazes in glassy polymers}

\author{J\"org Rottler}
\email{rottler@pha.jhu.edu}
\author{Mark~O.~Robbins}
\affiliation{Department of Physics and Astronomy, The Johns Hopkins
University, 3400 N.~Charles Street, Baltimore, Maryland 21218}

\date{\today}

\begin{abstract} 
We report on an extensive study of craze formation in glassy
polymers. Molecular dynamics simulations of a coarse-grained
bead-spring model were employed to investigate the molecular level
processes during craze nucleation, widening, and breakdown for a wide
range of temperature, polymer chain length $N$, entanglement length
$N_e$ and strength of adhesive interactions between polymer
chains. Craze widening proceeds via a fibril-drawing process at
constant drawing stress. The extension ratio is determined by the
entanglement length, and the characteristic length of stretched chain
segments in the polymer craze is $N_e/3$. In the craze, tension is
mostly carried by the covalent backbone bonds, and the force
distribution develops an exponential tail at large tensile forces. The
failure mode of crazes changes from disentanglement to scission for
$N/N_e\sim 10$, and breakdown through scission is governed by large
stress fluctuations. The simulations also reveal inconsistencies with
previous theoretical models of craze widening that were based on
continuum level hydrodynamics.
\end{abstract}


\maketitle

\section{Introduction}
\begin{figure}[t]
\begin{center}
\includegraphics[width=7cm]{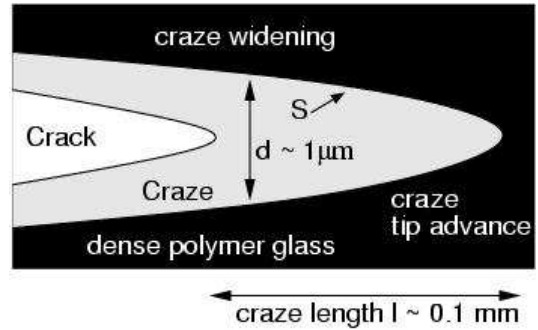}
\caption[Craze fracture of glassy polymers]{\label{geom-fig}Craze
fracture of glassy polymers. The craze is a deformed region (shaded)
that grows in width and length under an applied vertical stress
$S$. Its density is reduced with respect to the undeformed polymer by
a constant extension ratio $\lambda$. Characteristic values for width
$d$ and length $l$ are indicated. During growth, $S$ acts
perpendicular to the sharp interface between undeformed polymer and
craze. Also shown is an advancing crack tip from the left that breaks
the craze. Small representative volumes of each region are studied
with molecular simulations.}
\end{center}
\end{figure}

The failure of glassy polymers such as polystyrene (PS) or
polymethylmethacrylate (PMMA) under external stresses occurs either
through shear deformation or through crazing
\cite{Haward1997,Ward1983}. While shear yielding occurs essentially at
constant volume, crazing has a strong dilational component, and the
volume of the material increases to several times its original value
before catastrophic fracture occurs. Crazing is a failure mechanism
unique to entangled polymeric materials and usually precedes a crack
tip (see Fig.~\ref{geom-fig}). The fundamental and technological
importance of crazes is that they are in part responsible for the
large fracture energy $G_c$ of polymer glasses
\cite{Wool1995,Pocius1997,Brown1991,Rottler2002a} that makes them
useful load-bearing materials. They control the crack tip advance and
require a large amount of energy dissipation up to the point of
catastrophic failure. Crazes can reach several $\mu m$ in width and
consist of an intriguing network of fibrils and voids that spans the
entire deformed region.

Despite the frequent appearance of crazes, there is still
comparatively little theoretical understanding about the conditions
and mechanisms of craze nucleation, growth and ultimate breakdown
\cite{Haward1997,Ward1983,Wool1995,Pocius1997,Krupenkin1999,Kramer1990}. In
this paper, we present an extensive set of nonequilibrium molecular
dynamics (MD) simulations that address these various phenomena.  In
this approach, polymers are modeled on a coarse-grained scale that
takes into account van-der-Waals (vdW) and covalent interactions
without specific reference to chemical detail. The effect of chain
length, temperature $T$, widening velocity $v$ and vdW interaction
strength on the craze structure can be studied over a wide range of
parameters.  The molecular simulations allow insight into microscopic
details not accessible to experiments and offer an opportunity to test
and develop theoretical models of crazing.

A fundamental limitation on molecular level treatments is of course
the finite system size. The largest volumes accessible at
present are $\sim 100 nm^3$, while the craze spans many $\mu m$. We
are thus limited to a study of craze widening in a small
representative region and cannot include e.g. the entire crack
tip. Craze tip advance processes \cite{Kramer1990} are beyond the
scope of the present work.

Several aspects of craze physics have already been adressed with
simulations in previous papers. Baljon and Robbins
\cite{Baljon1996,Baljon2001} demonstrated the importance of chain
length for the onset of craze growth. Rottler {\it et al.}  studied
the elastic properties and fracture stresses of fully evolved crazes
and used them in combination with linear fracture mechanics to
calculate the macroscopic fracture energy of glassy polymers that fail
by crazing \cite{Rottler2002a}. Rottler and Robbins also investigated
how polymer entanglements affect the craze structure on a microscopic
level and argued that they ``jam'' the expansion of the glass under
tension \cite{Rottler2002b}.

This paper extends the previous work and is organized as follows. In
Section \ref{reviewsec}, we briefly summarize the key experimental
observations and review existing theoretical models of
crazing. Section \ref{modelsec} gives the technical details of the
molecular models used in this study. We then analyze results for craze
nucleation (Section \ref{nucsec}), growth (Section \ref{growthsec}),
microstructure (Section \ref{structsec}), and failure (Section
\ref{breaksec}), and compare our findings to previous models and
experiments.  Final conclusions are offered in Section \ref{concsec}.

\section{Craze phenomenology and theory}
\label{reviewsec}
\subsection{Experiments}
Crazes have been studied experimentally for more than 30 years
\cite{Argon1977,ArgonHan1977,Lauterwasser1979,Kramer1983,Kramer1990}. The
techniques most commonly used to analyse the craze structure are
transmission electron microscopy (TEM)
\cite{Washiyama1992,Michler1986}, low angle electron diffraction
(LAED) and small angle x-ray scattering (SAXS)
\cite{Paredes1979,Brown1981,Brown1987}. Comprehesive reviews of
theoretical and experimental results have been presented by Kramer and
Berger \cite{Kramer1983,Kramer1990} and Creton {\it et al.}
\cite{Creton2001}.

The density in the undeformed polymer $\rho_i$ and craze $\rho_f$ is
obtained from TEM measurements.  The increase in volume during craze
formation, or extension ratio $\lambda\equiv \rho_i/\rho_f$, is found to have
a characteristic value for a given polymer that is independent of
molecular weight. Typical values of $\lambda$ for different polymers
range from two to seven. 

Real space images of the craze show that the polymers are bundled into
fibrils that merge and split to form an intricate network. The fibrils
are highly aligned with the applied tensile stress and vary in
diameter and length. However, this complex structure is normally
idealized as a set of uniform vertical cylinders connected by short
cross-tie fibrils \cite{Kramer1990}.  The characteristic fibril
diameter $\langle D\rangle$ and separation $\langle D_0\rangle$ (see
Fig.~\ref{capmodel-fig}) are then determined from a Porod analysis of
scattering experiments (see Section \ref{growthsec}).  Measured values
range between $3-30$ nm for $\langle D\rangle$ and $20-50$ nm for
$\langle D_0\rangle$
\cite{Kramer1983,Kramer1990,Salomons1999a,Salomons1999b}.  For
example, for polystyrene one obtains $\langle D\rangle\sim 6$nm and
$\langle D_0\rangle\sim 20$nm \cite{Haward1997}.

Nucleation of crazes \cite{ArgonHan1977} occurs preferentially near
defects in the polymer.  These produce large local tensile stresses
that lead to the formation of microvoids that evolve into a craze.
Once nucleated, the craze grows in length and width (see
Fig.~\ref{geom-fig}).  It is well established \cite{Kramer1990} that
the craze widens by drawing material from the dense polymer into new
fibrils.  This deformation is confined to a narrow active zone at the
interface between the dense polymer and craze.  The width of the
active zone $h$ (see also Fig.~\ref{capmodel-fig}) is usually between
$\langle D\rangle$ and $\langle D_0\rangle$.  Craze widening is a
steady-state process, in which a constant ``drawing stress'' $S$
ranging between $20$ and $100$MPa is applied.  Typical experimental
values are 35 MPa (polystyrene) \cite{Kramer1983} and 70 MPa
(polymethymethacrylate) \cite{Brown1990}. The value of $S$ is of the
same order as the shear yield stress of the polymer, and is found to
increase with the entanglement density.

\subsection{Theory}
A theory of crazing has to explain the molecular origin of the craze
structure and the interdependencies of the various quantities measured
in experiments. Despite a wealth of experimental data on crazing,
there is currently no theoretical description that addresses all
aspects of craze physics. The following models have been proposed to
explain the extension ratio $\lambda$ and the relationship between
fibril spacing $\langle D_0 \rangle$ and drawing stress $S$.
 
\subsubsection{The extension ratio $\lambda$}
\label{lambda-subsubsec}
The extension ratio $\lambda$ has been successfully explained by a
simple scaling argument, that relates $\lambda$ to the microscopic
entanglement network in the polymer glass. Entanglements arise in
dense polymeric systems from the topological constraints that the
chains impose upon each other. The mobility of the chains is greatly
restricted, because they cannot pass through each other. The starting
point for the present argument is the assumption that the glass
inherits these entanglements from the melt, where an entanglement
molecular weight is given by the plateau modulus under shear,
$G_N^{(0)}$:
\begin{equation}
M_e=\rho 4RT/5G_N^{(0)}.
\label{nedef-eq}
\end{equation}
This result can be derived from the microscopic tube model
\cite{Doi1986}, which relates the rheological response of the polymer
melt to the deformation of a tube to which the polymer chain is
confined. With repeat units of weight $M_0$, one can define a typical
number of steps $N_e=M_e/M_0$ (entanglement length) between
entanglements along the polymer backbone. 

These entanglements are assumed to act like permanent chemical
crosslinks during crazing, which implies that the expansion ends when
segments of length $N_e$ are fully stretched. The initial separation
of entanglement points is $d_i=(l_p l_0 N_e)^{1/2}$, according to
standard random walk (RW) scaling, where $l_0$ is an elementary step
length and $l_p$ the persistence length. The length of this
segment rises from $d_i$ to a maximum final length $d_f=\lambda_{\rm
max} d_i=N_el_0$, and thus
\begin{equation}
\label{lambda-eq}
\lambda_{\rm max}=(N_el_0/l_p)^{1/2}. 
\end{equation}
Experimentally, Eq.~(\ref{lambda-eq}) is well confirmed, but Section
\ref{entang-subsec} shows that the picture motivating this expression
is oversimplified.

\subsubsection{The drawing stress $S$}
\begin{figure}[t]
\begin{center}
\includegraphics[width=7cm]{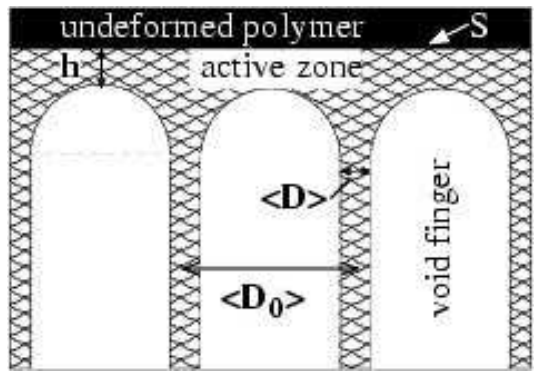}
\caption[Surface tension model of craze
widening]{\label{capmodel-fig}Surface tension model of craze
widening. Void fingers with characteristic spacing $\langle
D_0\rangle$ propagate into a strain-softened layer of polymer fluid of
width $h$, leaving behind fibrils of a characteristic diameter
$\langle D\rangle$. The externally applied stress $S$ acts
perpendicular to the fluid-glass interface. The characteristic radius
of the finger caps is on the order of $\langle D_0\rangle/2$. (See
Refs.~\protect\cite{Kramer1983,Kramer1990} for an analogous
figure.)}
\end{center}
\end{figure}

The value of the drawing stress $S$ has traditionally been related to
the craze microstructure ($\langle D\rangle$, $\langle D_0\rangle$)
via capillary models \cite{Kramer1983,Kramer1990}. In these models,
the polymer in the active zone is treated as a viscous fluid with a
surface tension $\Gamma$ and a viscosity $\eta$. Figure
\ref{capmodel-fig} shows an idealized picture of the craze geometry,
where craze formation is modelled as the propagation of void fingers
with a characteristic spacing $\langle D_0\rangle$ into the
strain-softened fluid. The applied stress $S$ required to advance the
interface has a dissipative contribution arising from a suitable flow
law (e.g. power-law fluid) and an energy penalty contribution due to
the surface tension. The tension is $S$ in the polymer glass and the
Laplace pressure $2\Gamma/(\langle D_0\rangle/2)$ at the ceiling of
the finger, where $\langle D_0\rangle/2$ is the characteristic radius
of curvature (see Fig.~\ref{capmodel-fig}). By estimating the width of
the active zone as $h\sim \langle D_0\rangle/2$, Kramer calculated a
stress gradient between glassy polymer and the finger void ceiling,
\begin{equation}
\nabla\sigma\sim \frac{\Delta \sigma}{h}\sim \frac{S-4\Gamma/\langle D_0\rangle}{\langle D_0\rangle/2}.
\end{equation}
Since $\nabla\sigma$ is proportional to the interface velocity, he
then predicted that the system will select a value of
\begin{equation}
\label{kramer-eq}
\langle D_0\rangle \sim 8\Gamma/S,
\end{equation}
which maximizes the stress gradient between finger ceiling and bulk
polymer and thus will lead to the fastest propagation velocity of the
fingers.

More recently, Krupenkin and Fredrickson \cite{Krupenkin1999} have
formulated a theory of craze widening that is similar in spirit to
Kramer's arguments and also equates the craze widening stress with a
viscous and a surface tension contribution. However, these authors
suggest a different interpretation of $\Gamma$. They introduce an
effective surface tension that begins to rise above the vdW value when
the finger radius rises above the rms spacing between entanglement
lengths $d_i$. This ansatz is motivated by the idea that expanding the
random walk between entanglements generates an additional energy
penalty. An upper bound to $\Gamma$ is provided by the energy required
for chain breaking, which sets in once the finger radius exceeds the
maximum elongation between entanglement points, $l_0N_e$.  By
minimizing the finger propagation stress, they conclude that the
fibril spacing will always be
\begin{equation}
D_0\sim d_i,
\end{equation}
independent of surface tension. In their model, the fibril spacing is
determined exclusively by the entanglement network.

\section{Simulations and Molecular Models}
\label{modelsec}
We study craze formation by performing molecular dynamics simulations
of a standard coarse-grained polymer model \cite{Puetz2000}, where
each linear polymer contains $N$ spherical beads of mass $m$. Models
of this kind have a long tradition in polymer research and have
verified theories of polymer dynamics \cite{Doi1986} in the melt. They
have recently been employed by other researchers to study failure in
network polymer adhesives \cite{Stevens2001} and end-grafted polymer
chains between surfaces \cite{Sides2001}. 

In this bead-spring model, van der Waals interactions between beads
separated by a distance $r$ are modeled with a truncated Lennard-Jones
potential:
\begin{equation}
V_{\rm
LJ}(r)=4u_0\left[(a/r)^{12}-(a/r)^{6}-(a/r_c)^{12}+(a/r_c)^{6}\right]
\label{LJ-pot-eq}
\end{equation}
for $r\le r_c$, where $u_0\sim 20-40$ meV and $a\sim 0.8-1.5$ nm are
characteristic energy and length scales \cite{Kremer1990}. A simple
analytic potential \cite{Sides2001}
\begin{equation}
V_{\rm br}(r)=-k_1(r-R_0)^3(r-R_1)
\label{cov-pot-eq}
\end{equation}
is used for covalent bonds between adjacent beads along the chain. The
form of this potential was chosen to allow for covalent bond breaking,
which is not possible with other standard bond potentials such as the
popular FENE potential \cite{Puetz2000}. Bonds are permanently broken
when $r$ exceeds $R_0=1.5\,a$. The constant $R_1=0.7575a$ was chosen
to set the equilibrium bond length $l_0=0.96\,a$, which is the
``canonical'' value for the bead-spring model with the FENE potential
\cite{Puetz2000}. This allows us to use results from previous studies,
most importantly the entanglement length.  The constant $k_1$
determines the ratio of the forces at which covalent and van der Waals
bonds break.  We find that this ratio is the only important parameter
in the covalent potential and set it to 100 based on data for real
polymers \cite{Sides2001,Stevens2001}, which implies
$k_1=2351u_0/a^4$. Tests with other analytical forms of the bond
potential showed no appreciable impact on our results as long as the
bonds break before the chains can pass through each other.

In order to vary the entanglement length, we include a bond-bending
potential \cite{Sides2001,Faller2000}
\begin{equation}
\label{angle-eq}
V_{B}=b\sum_{i=2}^{N-1}\left(1-\frac{(\vec{r}_{i-1}-\vec{r}_{i})\cdot
(\vec{r}_{i}-\vec{r}_{i+1})}{|(\vec{r}_{i-1}-\vec{r}_{i})||(\vec{r}_{i}-
\vec{r}_{i+1})|}\right)
\end{equation}
that stiffens the chain locally and increases the radius of
gyration. Here, $\vec{r}_{i}$ denotes the position of the $i$th bead
along the chain, and $b$ characterizes the stiffness. We will consider
two cases here referred to as flexible $(b=0)$ and semiflexible
$(b=1.5u_0)$ polymers. The corresponding entanglement lengths are
$N_e^{\rm fl}\approx 70$ and $N_e^{\rm sfl} \approx 30$ beads,
respectively \cite{Kremer1990,Puetz2000,Faller2000}.

We consider three temperatures $T=0.01\, u_0/k_B$, $T=0.1\, u_0/k_B$
and $T=0.3\,u_0/k_B$, where the last temperature is close to the glass
transition temperature. The amount of adhesive interaction between
beads is varied by changing the range $r_c$ of the LJ potential from
$1.5 a$ to $2.2 a$.

The equations of motion are solved using the velocity Verlet algorithm
with a timestep of $dt=0.0075\,\tau_{\rm LJ}$, where $\tau_{\rm
LJ}=\sqrt{ma^2/u_0}$ is the characteristic time given by the LJ energy
and length scales. Periodic boundary conditions are employed in all
directions to eliminate edge effects. The temperature is controlled
with a Nos\'e-Hoover thermostat (thermostat rate $1\,\tau_{\rm
LJ}^{-1}$), and the thermostat is only employed perpendicular to the
direction of craze growth. Simulations with a Langevin thermostat
showed no appreciable difference between the two methods.

In all simulations of crazing, an initial isotropic state in a cubic
simulation cell of edge length $L$ is created using standard techniques
\cite{Kremer1990}. Polymer chains are constructed as ideal RWs with a
suitably chosen persistence length $l_p$. $l_p$ is fixed by matching
the radius of gyration of the chains to the equilibrium value in the
melt, and the values are $l_p^{\rm fl}=1.65a$ and $l_p^{\rm sfl}=2.7a$
for flexible and semiflexible chains, respectively. Subsequently, the
interaction potentials are imposed and the system is cooled at
constant volume from a melt temperature $T_m=1.3 u_0/k_B$, to the
desired run temperature. 

All runs begin at zero hydrostatic pressure. Strains $\epsilon_{ii}$
are then imposed by rescaling the simulation box periods $L_i$ and all
particle coordinates proportionately \cite{Allen1987}. This allows
arbitrary stress states to be studied in Section \ref{nucsec}.

\section{Criteria for cavitation and craze nucleation}
\label{nucsec}
The loading conditions on the polymer glass determine whether it will
fail initially by shear yielding or the formation of voids and
cavities. In general, strong triaxial tensile stresses will favor
cavitation. Cavitation and crazing are closely related, because crazes
usually require the initial formation of microvoids
\cite{Kramer1990}. We therefore first address the initial failure of
the polymer glass through either shear yielding or cavitation, and
later discuss the formation of crazes.

The loading conditions that lead to shear yielding in many
experimental polymers \cite{Kody1997,Raghava1973} are most accurately
represented by the pressure-modified von Mises yield criterion. It is
formulated in terms of simple stress invariants, the hydrostatic
pressure $p=-(\sigma_1+\sigma_2+\sigma_3)/3$ and the deviatoric or
octahedral shear stress $\tau_{\rm
oct}=\left((\sigma_{1}-\sigma_{2})^2+(\sigma_{2}-\sigma_
{3})^2+(\sigma_{3}-\sigma_{1})^2\right)^{1/2}/3$, where the $\sigma_i$
denote the three principal stress components. The pressure-modified
von Mises criterion states that yield will occur at an octahedral
yield stress $\tau_{oct}^{y}$ given by
\begin{equation}
\label{pressure-eq}
\tau_{oct}^{y}=\tau_0+\alpha p,
\end{equation}
where $\tau_0$ is the yield stress at zero hydrostatic pressure and
$\alpha$ is a dimensionless constant. Its physical motivation is that
the elastic free energy stored in shear deformation is proportional to
$\tau_{\rm oct}^2$ and failure should occur when this energy exceeds a
threshold that rises slowly with $p$.

\begin{figure}[tb]
\begin{center}
\includegraphics[width=7cm]{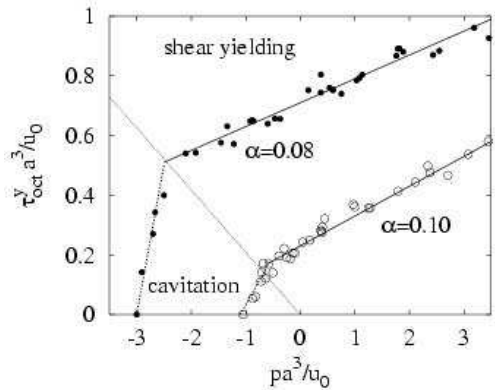}
\caption[Transition from shear yielding to
cavitation]{\label{shearcav-fig}Octahedral shear stress $\tau_{\rm
oct}^y$ at yield as a function of pressure $p$ at two different
temperatures $T=0.3u_0/k_B$ (open symbols) and $T=0.01u_0/k_B$ (filled
symbols).  The solid lines are fits to Eq.~(\ref{pressure-eq}) and the
dashed lines show the onset of cavitation. Values of $\alpha$ are
indicated for the two temperatures. Also drawn is a dotted line
through the transition points that separates the regions of shear and
cavitational failure. Here yield is associated with the strain where
$\tau_{\rm oct}$ peaks.}
\end{center}
\end{figure}

In ref.~\onlinecite{Rottler2001}, we examined a much larger range of
stress states than in previous experimental studies and showed that
the pressure-modified von Mises criterion provides a good description
of shear yield in our bead-spring model. Data for two extremal
temperatures are replotted in Fig.~\ref{shearcav-fig} along with solid
lines showing fits to Eq.~(\ref{pressure-eq}).  Shear yield was
observed to the right of the dot-dashed line, and these data points
follow Eq.~(\ref{pressure-eq}) quite accurately. To the left of the
line cavitation was observed.  The deviation from the von Mises fits
is very sharp, and $\tau_{\rm oct}$ quickly drops to zero. The values
of $\tau_{\rm oct}^c$ where cavitation occurs are well described by a
straight lines
\begin{equation}
\label{av-eq}
\tau_{oct}^{c}=\tau_0^c+\alpha^c p
\end{equation}
with new constants $\tau_0^c$ and $\alpha^c$. This new ``cavitation
criterion'' can be motivated in analogy to the von Mises criterion by
assuming that the elastic free energy $F_{V}$ associated with volume
changes must reach a critical value for cavitation to occur. $F_{V}$
is proportional to $p^2$, which gives a criterion of the form
$p=p_0$. One can then assume that shear components in the stress
tensor aid cavitation in a linear fashion, i.e. $p=p_0+\tau_{\rm
oct}/\alpha_c$, which can be rearranged to give Eq.~(\ref{av-eq}) with
$\tau_0^c=\alpha_cp_0$.

No clear experimental consensus exists about the stress state required
for crazing, partly because of the importance of surface defects in
nucleating crazes.  However, several criteria for craze nucleation
were proposed almost 30 years ago. They all try to take into
account the critical role of tensile stress components. Sternstein {\it
et al.}  \cite{Courtney1990} suggested a craze yield criterion of the
form
\begin{equation}
\label{sternstein-eq}
\tau_{\rm max}\equiv\frac{1}{2}|\sigma_i-\sigma_j|_{\rm max}=A+B/p,
\end{equation}
where $A$ and $B$ are constants that depend on temperature.  With
respect to our criterion Eq.~(\ref{av-eq}), $p$ has been replaced by
$1/p$ and $\tau_{\rm oct}$ by the largest difference between any two
stress components.  Bowden and Oxborough \cite{Haward1997} formulated
a similar criterion, where $\tau_{\rm max}$ is replaced by
$\sigma_1-\nu\sigma_2-\nu\sigma_3$ and $\nu$ is Poisson's ratio for
the polymer glass. This expression is another possibility to describe
the shear components of the stress state, and it reduces to $\tau_{\rm
max}$ when $\nu=1/2$ and $\sigma_2=\sigma_3$. The Sternstein and
Bowden and Oxborough expressions could in principle also be fitted to
the rather narrow range of pressure in Fig.~\ref{shearcav-fig} where
cavitation occurs. However, we are unaware of a convincing physical
motivation for the $1/p$ term, which leads to obvious analytical
problems at small $p$. In addition, the experimental results that
motivated Eq.~(\ref{sternstein-eq}) are sensitive to surface defects
\cite{ArgonHan1977}.

The above considerations pertain to the {\em initial} mode of failure
of the polymer glass at strains typically less than 10\%. However,
crazing is a large strain deformation with strains of several hundred
percent. Although we find voiding to be a neccessary precursor to
crazing, it is not guaranteed that a loading state that leads to
cavitational failure according to Eq.~(\ref{av-eq}) will ultimately
produce stable crazes. Likewise, we have observed that an initial
failure through shear deformation can still lead to later void
formation and crazing. One should thus strictly call Eq.~(\ref{av-eq})
a cavitation failure criterion and not a craze yielding criterion.

\section{Growth of Crazes}
\label{growthsec}
In order to induce crazing, we enforce cavitation by expanding the
periodic simulation box in the z-direction at constant velocity while
maintaining the simulation box periods in the perpendicular $x-y$
plane. This leads to an initial stress state where all three principal
stresses are tensile. The initial voids formed during cavitation
expand upon further straining, but their growth rapidly becomes
arrested \cite{Baljon2001}. Instead of forming new voids, additional
material is drawn out of the uncavitated polymer, and stable craze
growth occurs. In our simulations, growth continues until all material
in the simulation box is converted into the craze.

\subsection{Images of crazes}

\begin{figure}[t]
\begin{center}
\includegraphics[width=8cm]{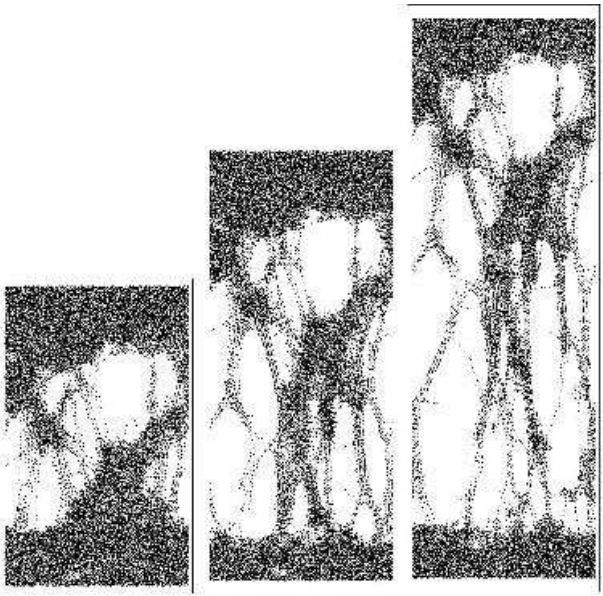}
\caption[Snapshots of crazes for flexible chains with $T=0.1 u_0/k_b$
and $r_c=1.5 a$]{\label{seq1-fig}Three snapshots of craze growth for
flexible chains with $T=0.1 u_0/k_B$ and $r_c=1.5 a$. The total system
contains 262144 beads, but only slices of thickness $10 a$ normal to
the page are shown in order to resolve the fine structure. The lateral
dimension of each slice is $64a$ and the vertical direction is to
scale. Each dot represents one Lennard-Jones bead.}
\end{center}
\end{figure}

\begin{figure}[t]
\begin{center}
\includegraphics[width=8cm]{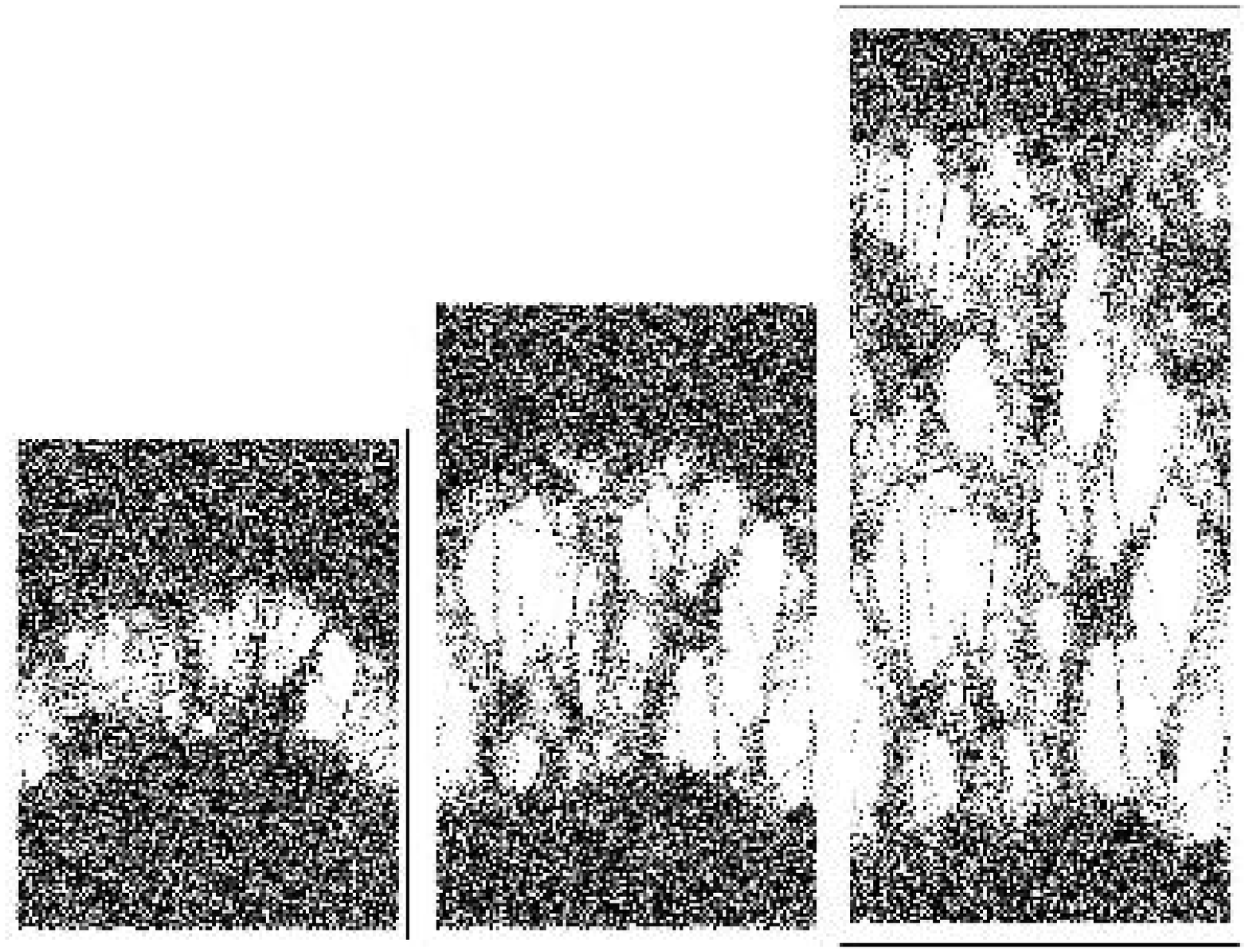}
\caption[Snapshots of crazes for semiflexible chains with $T=0.1
\,u_0/k_b$, $r_c=1.5\, a$]{\label{seq2-fig}Three snapshots of craze growth
for semiflexible chains with $T=0.1\, u_0/k_B$, $r_c=1.5\, a$, and 262144
beads.}
\end{center}
\end{figure}

\begin{figure}[t]
\begin{center}
\includegraphics[width=8cm]{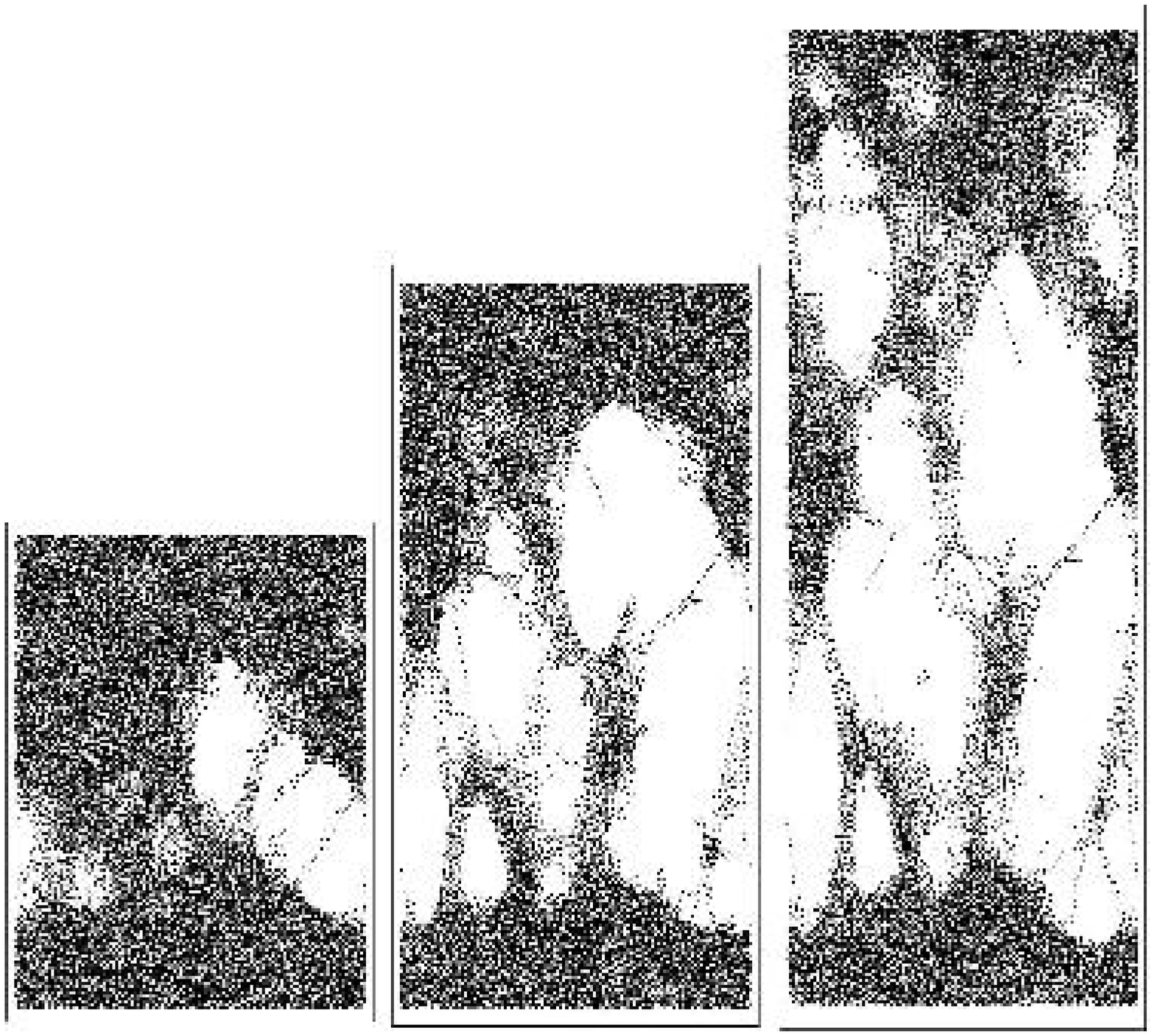}
\caption[Snapshots of crazes for semiflexible chains with $T=0.3\,
u_0/k_b$, $r_c=2.2\, a$]{\label{seq3-fig}Three snapshots of craze growth
for semiflexible chains with $T=0.3\, u_0/k_B$, $r_c=2.2\, a$, and 262144
beads.}
\end{center}
\end{figure}

A good impression of the crazing process can be obtained by inspecting
the snapshots of the simulation cell shown in Figs.~\ref{seq1-fig} -
\ref{seq3-fig}. Each slice has a lateral width of $64\,a$, and three
different strains are shown. In all images, the chain length
$N=512$. Previous studies \cite{Baljon2001} had shown that $N$ has to
be twice the entanglement length or greater in order to form stable
crazes. For shorter chains, the material cavitates, but then rapidly
fails due to chain pullout. In the following, we only consider chains
with $N\ge 2N_e$.

Note first that in all cases, there is a sharp interface between dense
polymer and crazed material. This narrow 'active zone' is one of the
key features of craze phenomenology found in experiment. In the craze,
the polymer chains have merged into fibrils that are strongly aligned.
However, the structure is quite complicated, as there are many lateral
connections between fibers.

One can also observe that the fine structure of the crazes in the
three sequences varies greatly. Fig.~\ref{seq1-fig} with flexible
chains at the low temperature of $T=0.1 u_0/k_B$ and the weak adhesive
interaction (cutoff distance $r_c=1.5 a$) shows many thin fibrils,
whereas the fibrils in Fig.~\ref{seq3-fig} at the higher temperature
of $T=0.3 u_0/k_B$ and the stronger adhesive interaction $r_c=2.2 a$
are much thicker in diameter. These trends are not surprising, because
increased chain mobility at higher temperatures and stronger adhesive
interactions should drive the system to larger fibril diameters, which
minimize the surface area.

\subsection{The drawing process and stress-strain curves}
\begin{figure}[t]
\begin{center}
\includegraphics[width=8cm]{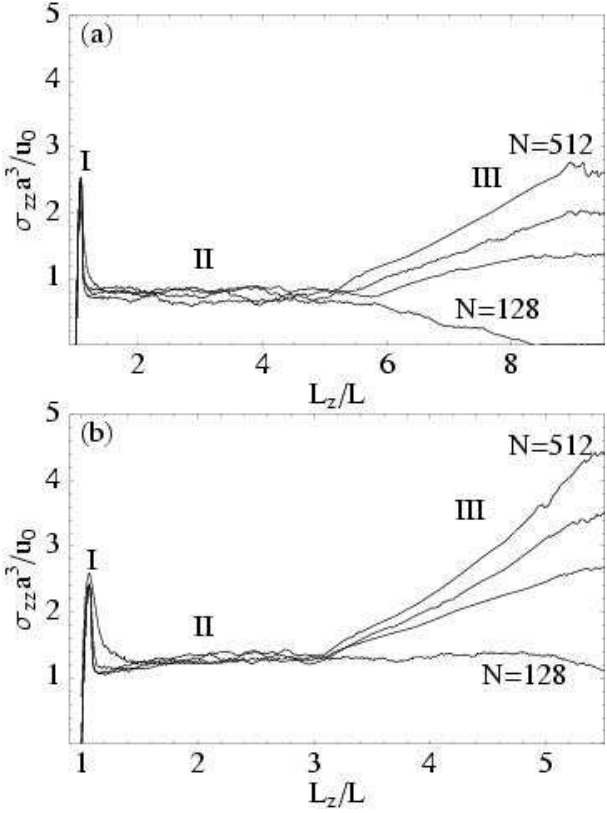}
\caption[Stress-strain curves during craze formation and
breakdown]{Stress $\sigma_{zz}$ in the widening direction during craze
growth at $T=0.1\, u_0/k_B$, $r_c=1.\,5 a$ for (a) flexible and (b)
semiflexible chains of length $N=128$, $N=256, N=384$, and $N=512$. Three
characteristic regimes of (I) cavity nucleation, (II) craze growth and (III)
craze failure are also indicated. The two perpendicular stress
components $\sigma_{xx}$ and $\sigma_{yy}$ also peak at cavitation (see
text), but then rapidly drop to zero. Qualitatively identical curves
are obtained at other values of $T$ and $r_c$.}
\label{stressstrain-fig}
\end{center}
\end{figure}

A second characteristic feature of craze growth is that deformation
occurs at a constant plateau or drawing stress $S$. This plateau can
be easily identified in the stress-strain curves shown in
Fig.~\ref{stressstrain-fig}. The curves can be separated into three
different regimes. In regime I, the stress rises to a peak of $\sim
2.6 u_0/a^3$ and then drops when the polymer yields by cavitation.
Following cavitation, the stress rapidly relaxes and remains at the
plateau value $S$ in regime II, the growth regime. Regime II is much
shorter in the semiflexible case Fig.~\ref{stressstrain-fig}(b) than
in the flexible case Fig.~\ref{stressstrain-fig}(a) (note different
lateral scales). Regime II ends when the strain $L_z/L$ reaches the
extension ratio $\lambda$. At this point, all the material in the
simulation cell has been converted into the craze, and additional
deformation strains the entire craze uniformly. As a consequence, the
stress rises again in regime III. This regime finally ends in
catastrophic failure either through chain disentanglement or chain
scission (see Section
\ref{breaksec}).

Note first that neither the peak stress at cavitation nor the value of
$S$ depends on the chain length $N$.  The curves for different $N$ in
Fig.~\ref{stressstrain-fig} only split apart after completion of craze
growth when $L_z/L$ reaches $\lambda$ and the entire craze is
strained. Baljon and Robbins \cite{Baljon2001} showed that the peak
stress remained constant for much shorter chains, but that regime II
only appeared when $N$ was $2N_e$ or longer. Another important fact
to note is that $S$ is independent of system size. For example, values
of $S$ in systems ranging between 32768 and 1048576 beads are the same
within a few percent. The biggest change with increasing system size
is that temporal fluctuations in $S$ decrease.

\begin{figure}[t]
\begin{center}
\includegraphics[width=8cm]{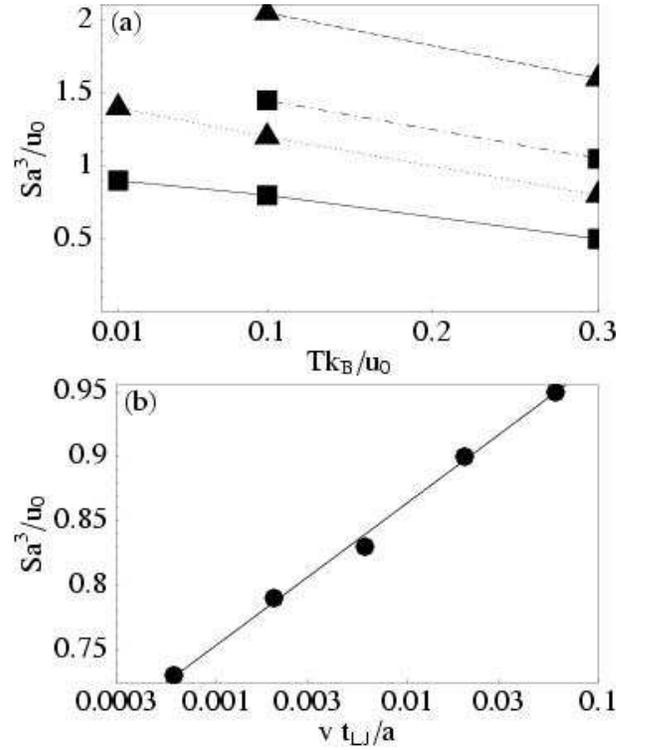}
\caption[Trends of $S$ with $T$ and $r_c$]{\label{Strends-fig}(a)
Trends of $S$ with $T$ and $r_c$ at $v=0.06 a/\tau_{\rm LJ}$ for
flexible ($\blacksquare$) and semiflexible ($\blacktriangle$) chains
and $r_c=1.5a$ (lower curves) and $r_c=2.2a$ (upper curves). (b)
Velocity dependence of $S$ for flexible chains at $T=0.1 u_0/k_B$. The
straight line is a fit to a logarithmic velocity dependence,
$S=1.085\,u_0/k_B+0.048\,u_0/k_B\ln{v}$. Uncertainties are comparable
to symbol sizes.}
\end{center}
\end{figure}

In Fig.~\ref{Strends-fig}(a), we analyze trends of $S$ with $T$ and
$r_c$. The drawing stress decreases linearly with increasing
temperature and increases with increasing adhesive interactions
(i.e. increasing $r_c$). Fig.~\ref{Strends-fig} (b) shows that $S$
varies logarithmically with the widening velocity $v$ over two orders
of magnitude, which is indicative of a thermally activated
process. For the subsequent figures, we choose $v=0.06 a/\tau_{\rm
LJ}$, which is at the upper end of the logarithmic regime
\cite{Baljon2001}. Similar behavior is also found for the shear yield
stress of glassy polymers \cite{Rottler2001,Rottler2003}.

\subsection{Crazing under plane stress conditions}
\begin{figure}[t]
\begin{center}
\includegraphics[width=8cm]{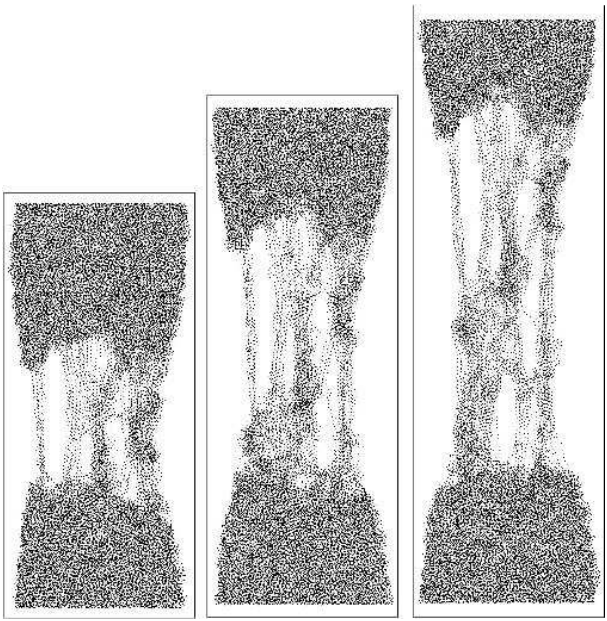}
\caption[Snapshots of crazes with a free
interface]{\label{biaxcraze-fig} Cross-sections through a craze with a
free interface at $T=0.1\, u_0/k_B$, $r_c=1.5\, a$, and 262144
beads. Periodic boundary conditions were maintained in the direction
into the plane. The location of initial cavitation was constrained by
placing repulsive beads in the center plane at $z=L_z/2$. The lateral
dimension is $47\,a$ and the vertical dimension is to scale.}
\end{center}
\end{figure}

The results of Section \ref{nucsec} show that cavitation only occurs
when all three principal stresses are tensile.  Many experimental
crazes grow in a thin film geometry under plane stress
conditions. However, in these experiments the craze is often
prenucleated or nucleates near a defect \cite{ArgonHan1977}. This
situation can also be mimicked in our simulations. To this end, the
periodic boundary conditions in the $x$ direction were replaced
with free boundaries, so that the solid is free to relax in that
direction. Initial failure is now nucleated by placing 1000 purely
repulsive LJ beads in the center plane of the simulation cell located
at $z=L_z/2$ \footnote{This is only done in
Fig.~\ref{biaxcraze-fig}. All other simulations use fully 3D periodic
boundary conditions without repulsive beads.}. This weakens the solid
locally and constrains the location of initial failure, while not
affecting subsequent craze growth.

Fig.~\ref{biaxcraze-fig} shows three snapshots of a craze in this
geometry. As in experiments, necking is observed at the craze-bulk
interface. Although $\sigma_{\rm xx}$ vanishes in the rest of the
film, the neck produces strong tensile stresses in all three
directions in the active zone. The craze grows in the same fashion as
in the simulations with 3D periodic boundary conditions. Since the
latter yield better statistics for the craze structure, we have
focused on this methodology for our analysis.

\subsection{The extension ratio}
\label{entang-subsec}
\begin{figure}[t]
\begin{center}
\includegraphics[width=8cm]{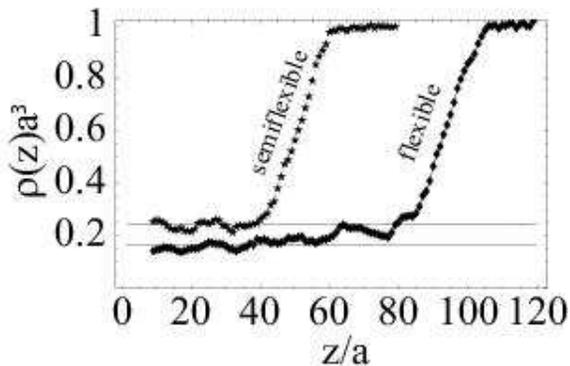}
\caption[Density profile through the active
zone]{\label{densitydrop-fig}Density profile through the active zone
for crazes with flexible chains ($N_e\approx 70$) and semiflexible
chains ($N_e\approx 30$). Horizontal lines indicate the average
density in the craze for the two cases. }
\end{center}
\end{figure}

The extension ratio $\lambda$ can be calculated from the average
densities of crazed and uncrazed material. Fig.~\ref{densitydrop-fig}
shows how the density drops from the initial value $\rho_i$ to
$\rho_f$ in the craze. As can be seen, $\rho_f$ is higher for the
semiflexible chains, which have a smaller value of $N_e\approx
30$. Remarkably, we find that $\lambda$ is a function of $N_e$ only
and decreases with decreasing $N_e$. For instance, while increasing
$T$ and $r_c$ produces dramatic coarsening of the fibril structure in
Fig.~\ref{seq3-fig} relative to Fig.~\ref{seq1-fig}, $\lambda$ is
unchanged. We obtain values of $\lambda_{\rm fl}=6.0 \pm 0.6$ and
$\lambda_{\rm sfl}=3.5 \pm 0.3$ independent of $N$, $T$, and adhesive
interaction strength.

\begin{figure}[t]
\begin{center}
\includegraphics[width=8cm]{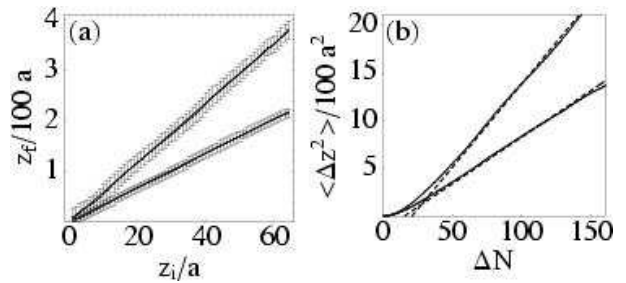}
\caption[Analysis of bead positions during crazing:
z-direction]{\label{affinedisp-fig}(a) Final bead heights $z_f$ as a
function of initial heights $z_i$ for flexible (large slope) and
semiflexible (small slope) chains ($T=0.1\,\epsilon/k_B$,
$r_c=1.5\,a$). Averages were calculated over z-intervals of width
$a$. Straight lines have slope $\lambda=5.9$ and $\lambda=3.5$,
respectively. Error bars represent a standard deviation from the
averages in each layer and are on the order of $19a$ (flexible) and
$9a$ (semiflexible). (b) Square of the height change $\Delta z$ as a
function of the number of covalent bonds $\Delta N$ between a bead and
the chain center. Dashed straight lines have slope $\lambda^2
l_pl_0/3$ with $\lambda$ from (a). Deviations from the RW scaling
occur in the vicinity of the chain ends (not shown). Other systems at
different $T,r_c$ and $N$ show the same results.}
\end{center}
\end{figure}

In order to understand the dependence of the macroscopic quantity
$\lambda$ on $N_e$, we analyze the structural changes in the polymer
glass during deformation on a microscopic level (see also
ref.~\cite{Rottler2002b}). Figure \ref{affinedisp-fig}(a) shows the
average final position of beads in the completely evolved craze as a
function of their initial positions along the direction of the
expansion (z-axis). The average was taken over all beads with initial
heights in a bin of width $1$a.  Although the strain rate is strongly
localized during the craze process, the ultimate displacement profile
is linear, $z_f=\lambda z_i$. 

To measure deviations from a purely affine (uniform) deformation, we
evaluated the rms variation $\delta z$ in $z_f$ for beads in each
bin. This quantity is indicated by error bars in
Fig.~\ref{affinedisp-fig}(a). Note that the variation in each bin is
very reproducible. We find that $\delta z$ is nearly independent of
$T$ and $r_c$ and has values on the order of 19a and 9a for flexible
and semiflexible chains, respectively.

\begin{figure}[tb]
\begin{center}
\includegraphics[width=8cm]{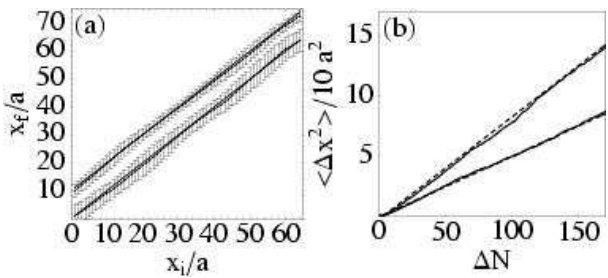}
\caption[Analysis of bead positions during crazing:
x-direction]{\label{affinedispx-fig} Analysis of bead positions
analogous to the previous figure (same systems), but for the
x-positions. No strain is applied in this direction, and the straight
lines in panel (a) have slope one. The curves for the semiflexible
chains in (a) were displaced vertically upward by 10a to avoid
overlap.  Error bars represent a standard deviation $\delta x$ from
the averages, and are on the order of $3.7a$ (flexible) and $2.2a$
(semiflexible). (b) Bead displacements as a function of distance from
the center in bond lengths, $\Delta N$, along the chain. Dashed lines
have slope $l_pl_0/3$.}
\end{center}
\end{figure}

Since no strain is applied in the perpendicular x and y directions,
one would assume that there is on average no displacement in these
directions. That this is indeed the case is shown in
Fig.~\ref{affinedispx-fig}, which repeats the analysis of
Fig.~\ref{affinedisp-fig} for the x-direction. Average final bead
positions are identical to initial positions, but there are lateral
variations $\delta x$ that are indicated by error bars. These lateral
displacements allow chains to gather in fibrils at the initial density
to minimize surface area. Unlike the vertical displacements $\delta
z$, these lateral displacements depend strongly on $T$ and $r_c$.  For
example, $\delta x \sim 2.5$a for the fine structure shown in
Fig.~\ref{seq1-fig}, where many thin fibrils can be seen, while
$\delta x \sim 5.6$a for the much coarser structure of
Fig.~\ref{seq3-fig}. In general, $\delta x$ correlates with the
spacing between fibrils as discussed in Section \ref{structsec} and
is less than $d_i$.  Krupenkin and Fredrickson
\cite{Krupenkin1999} suggested that $d_i$ provides an upper bound for 
the lateral chain deformations.

We now examine changes in the conformation of individual chains.  In
the initial state, the polymer chains exhibit an ideal random walk
(RW) structure inherited from the melt. The average end-to-end vector
$\langle\bf{R}^2\rangle$ thus scales with the number of covalent bonds
connecting two beads $\Delta N$ as $\langle{\bf
R}^2\rangle=l_pl_0\Delta N$. The component along each direction is
$1/3$ of that value since the initial state is
isotropic. Fig.~\ref{affinedispx-fig}(b) shows this initial scaling
behavior for $\langle\Delta x^2\rangle$ (dashed line) and that
$\langle\Delta x^2\rangle$ is not affected by crazing (solid
line). 

After an affine deformation by $\lambda$ along $z$, one would have an
anisotropic RW with no change in $\Delta x$ or $\Delta y$, but
$\langle\Delta z^2\rangle=\lambda^2 l_pl_0\Delta
N/3$. Fig.~\ref{affinedisp-fig}(b) shows the actual behavior (solid
lines) of $\langle\Delta z^2\rangle$ in the craze. At large scales, it
exhibits the expected scaling for an affine deformation (dashed
lines). However, the separation between beads is fixed by the length
of the covalent bonds, so the deformation of individual polymers along
$z$ cannot be purely affine. At small scales, the linear scaling
behavior of $\langle\Delta z^2\rangle$ crosses over into a quadratic
behavior, which indicates that the polymer has been pulled taut on
this scale.  The typical number of beads in such a straight segment
$\tilde{N}_{\rm st}$ can be calculated by letting $(\tilde{N}_{\rm
st}l_0)^2=\langle\Delta z^2\rangle=\lambda^2l_pl_0N_{\rm st}/3$ at the
crossover point, which yields $\tilde{N}_{\rm
st}=\lambda^2l_p/3l_0$. Inserting the observed values of $\lambda$,
$l_p$, and $l_0$, we arrive at $\tilde{N}_{\rm st}^{\rm fl}=21\pm 4$
and $\tilde{N}_{\rm st}^{\rm sfl}=12\pm 2$, respectively. These values
are comparable to the values of $\delta z$ found in
Fig.~\ref{affinedisp-fig}(a). On this scale, the deformation is
non-affine.

\begin{figure}[t]
\begin{center}
\includegraphics[width=8cm]{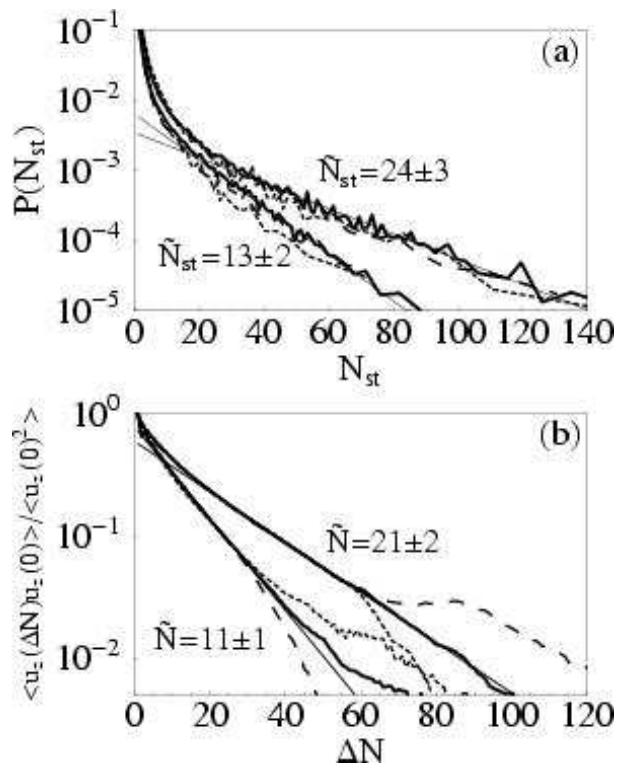}
\vspace{0.1cm}
\caption[Characteristic lengths of straight
segments]{\label{straightsegments-fig}(a) Probability distribution of
straight segments of length $N_{\rm st}$ for flexible and semiflexible
chains.  Thick lines correspond to simulations at
$T=0.1\,\epsilon/k_B$, $r_c=1.5\sigma, N=512$ with 1048576
beads. Dotted lines were obtained at $T=0.3\,\epsilon/k_B$,
$r_c=2.2\sigma, N=512$ with 262144 beads, and long dashed lines
correspond to $T=0.1\,\epsilon/k_B$, $r_c=1.5\sigma , N=256$ with
262144 beads. The straight lines show fits to $ \exp(-N_{\rm
st}/\tilde{N}_{\rm st}^{\rm (s)fl})$. (b) z-component of the bond-bond
correlation function for the same systems. Thin solid lines show
exponential fits with the indicated decay lengths.}
\end{center}
\end{figure}

The length of taut sections can also be determined by direct analysis
of the chain geometry. To this end, we calculate the angle between
every covalent bond and the z-axis and label a bond as pointing up
(down) if the angle is within $45^\circ$ of the z (-z) axis. We then
count the number $N_{\rm st}$ of consecutive up (down) steps. The
probability $P(N_{\rm st})$ of finding a straight segment containing
$N_{st}$ steps is shown in Fig.~\ref{straightsegments-fig}(a). For
both flexible and semiflexible chains, the distribution develops an
exponential tail. Like $\lambda$, this tail is independent of $N$,
$T$, and $r_c$.  The characteristic length scales that arise from
these tails are $\tilde{N}_{\rm st}^{\rm fl}\sim 21$ and
$\tilde{N}_{\rm st}^{\rm sfl}\sim 13$, in good agreement with the
prediction from the RW argument. Fig.~\ref{straightsegments-fig}(b)
shows that very similar length scales arise from an equivalent
analysis of the decay of the correlation function for the z-component
of successive bonds.

In section \ref{lambda-subsubsec}, we introduced the standard scaling
argument Eq.~(\ref{lambda-eq}) that relates extension ratio and
entanglement length, which has been verified experimentally with great
success.  In our cases, it predicts $\lambda_{\rm max}^{\rm fl}=6.5$
and $\lambda_{\rm max}^{\rm sfl}=3.5$, which agree with the observed
values of $\lambda$. However, the argument was motivated by the idea
that segments between entanglements become fully stretched and thus
appears to be at odds with the finding of an average straight segment
length of only $N_e/3$ rather than $N_e$. This discrepancy is resolved
by realizing that since the deformation is uniaxial, only the
projection of $d_i$ onto the z-axis, $d_i\cos(\Theta_z)$, is expanded,
where $\Theta_z$ is the angle between $d_i$ and the z-axis. The {\em
average} projection is thus only $1/\sqrt{3}$ of the total
length. Indeed, it was noted already in earlier work \cite{Kramer1983}
that, due to this geometric factor, $\lambda$ should be
$\sqrt{3}\lambda_{\rm max}$ for fully stretched chains. However, this
result is little cited since $\lambda\approx\lambda_{\rm max}$ in many
systems and, until our work, there was no reason to expect the length
of straight segments to be $N_e/3$.

The emergence of the length scale $N_e/3$ is a consequence of the
random nature of the entanglement mesh. Clearly, all strands would be
expanded simultaneously by the same factor in a regular mesh as
reported in a simulation study by Stevens \cite{Stevens2001}. By
contrast, in the polymer glass only the segments that are initially
aligned with the stretching direction become fully stretched. These
fully stretched segments are able to prevent further extension,
because the entanglements act like chemical crosslinks. Barsky and
Robbins have confirmed the equivalence between entanglements and
crosslinks by adding permanent crosslinks randomly to the system
\cite{Barsky2000}. The length between constraints then decreases from
$N_e$, and $\lambda_{\rm max}$ decreases accordingly. They found
$\lambda\approx\lambda_{\rm max}$ in all cases and that the {\em
average} stretched length $N_{\rm st}$ remains at $1/3$ of the distance
between constraints.

The success of the scaling argument Eq.~(\ref{lambda-eq}) and the
constancy of the extension ratio imply that there is no appreciable
loss of entanglements in our simulations during craze growth. Chains
do not disentangle once $N>2N_e$, and chain scission (see also section
\ref{breaksec}) is not observed during growth for any choice of
parameters in our model.
\begin{table}[tb]
\begin{center}
\caption[Dissipation during craze growth and covalent contribution to
the crazing stress $S$]{\label{edis-table}Dissipation during craze
growth and covalent contribution to the crazing stress $S$ for several
different systems of size 262144.}
\begin{tabular}{lccccccc}
 & T& $r_c$ & N & $\delta Q/\delta W$ & \% cov stress \\\hline
fl. & 0.1 & 1.5 & 256  & 0.88 & 87 \\ 
fl. & 0.1 & 1.5 & 512  & 0.88 & 88 \\
sfl. & 0.1 & 1.5 & 256  & 0.71 & 95 \\ 
sfl. & 0.1 & 1.5 & 512  & 0.67 & 97 \\
fl. & 0.1 & 2.2 & 512  & 0.92 & 69 \\ 
sfl. & 0.1 & 2.2 & 512  & 0.71 & 75 \\ 
fl. & 0.3 & 2.2 & 512  & 0.87 & 61 \\ 
sfl. & 0.3 & 2.2 & 512 & 0.78 & 67  
\end{tabular}
\end{center}
\end{table}

\subsection{Energy dissipation and stress transfer during crazing}
The work done in transforming a volume $dV$ of polymer into a craze is
$\delta W=S(\lambda-1)dV$. This work can either increase the potential
energy $dU$ or be dissipated as heat $\delta Q$. The division between
energy and heat is hard to determine experimentally, but simulations
with short chains found that both contributions were substantial
\cite{Baljon1996}. We have measured $\delta W$ and the energy change
directly in our simulations and calculated $\delta Q$ using the first
law of thermodynamics: $\delta Q=dU-\delta W$.  $dU$ can be calculated
directly from the bead positions and interaction potentials.  Table
\ref{edis-table} shows the fraction $\delta Q/\delta W$ of
dissipated total work for a number of large systems.  In all cases, a
large percentage, $\sim 80\%$, of the total work is dissipated, and
only $\sim 20\%$ is stored as potential energy.
\begin{figure}[t]
\begin{center}
\includegraphics[width=8cm]{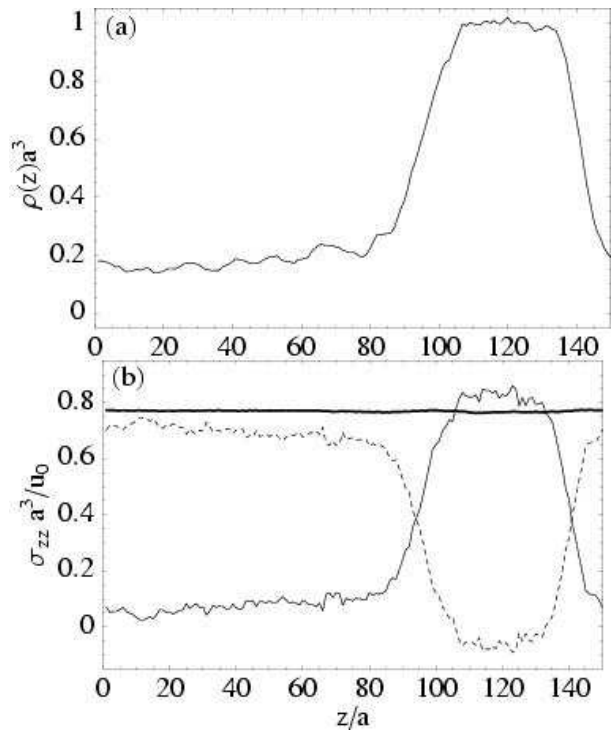}
\vspace{0.1cm}
\caption[Stress transfer during
crazing]{\label{stresscrazeuncr-fig}(a) Density profile through a
craze simulation at $T=0.1u_0/k_B$, $r_c=1.5a$ with flexible
chains. (b) vdW stress (solid line), covalent stress (dashed line) and
total stress $S$ (thick line) as a function of position along z. The
kinetic contribution to the stress is split evenly between the
covalent and vdW stress here and in Table \ref{edis-table}.}
\end{center}
\end{figure}
Since the craze drawing stress varies logarithmically with velocity
(see Fig.~\ref{Strends-fig}(b)), these percentages could change with
velocity. However, we find that $dU$ also decreases with decreasing
velocity, and there is no measurable change in the percentage of work
converted to heat over at least two orders of magnitude in velocity.

Stress in the craze can also be partitioned into two components that
originate either from van der Waals (LJ) interactions
Eq.~(\ref{LJ-pot-eq}) or from covalent interactions
Eq.~(\ref{cov-pot-eq}). The two contributions are very different in the
uncrazed and crazed material. In the undeformed polymer, the tensile
stress is mainly carried by the vdW bonds. This stress is transferred
in large part to the covalent bonds as they pass through the active
zone. Evidence for this is provided in Fig.~\ref{stresscrazeuncr-fig},
which shows the covalent and LJ contributions to the total stress as a
function of height in the widening direction. Panel (a) displays the
density profile in order to identify dense polymer regions (high
density) and craze regions (low density). In Panel (b) one observes
that in the dense region all the tension is carried by the van der
Waals bonds and the covalent bonds are under slight compression. In
the craze, between $60-95\%$ of the total stress (see Table
\ref{edis-table}) is carried by the covalent bonds, and the van der
Waals bonds only contribute a small fraction.

\subsection{Problems with the capillary models}
\label{S0-sec}
The results presented so far reveal serious difficulties with the
surface tension models discussed in the Introduction. The first
evidence of this comes from the observation that $S$ is independent of
system size. In our smallest simulations of lateral width $32 a$ the
simulation box only contains a few fibrils at $T=0.1 u_0/k_B,
r_c=2.2a$. If $S$ were controlled by Eq.~(\ref{kramer-eq}), one should
expect that the simulation box would need to contain a statistically
significant number of fibrils of spacing $\langle D_0\rangle$ for $S$
to reach its steady state value. However, the value of $S$ does not
fluctuate as the lateral dimensions are increased to $64 a$ or $128
a$.

The second and more severe problem concerns the distribution of stress
in the craze. The surface tension model assumes that {\em all} the
stress is carried at the interfaces of fibrils or in viscous stress in
the active zone. However, Table \ref{edis-table} and
Fig.~\ref{stresscrazeuncr-fig} show that almost all the stress in the
fibrils is carried by the covalent bonds, while the surface tension is
entirely associated with broken vdW bonds and small entropic
contributions.

In the following, we make an alternative proposition to relate craze
microstructure and drawing stress. This proposition is based on the
observation that the values of $S$ and $\lambda$ obtained from our
simulations obey the equality
\begin{equation}
\label{Slambda-eq}
S_{\rm fl}\lambda_{\rm fl}=S_{\rm sfl}\lambda_{\rm sfl}=S_0(T,r_c,v).
\end{equation}
This can be verified for each $T$ and $r_c$ using $\lambda_{\rm
fl}=6.0$, $\lambda_{\rm sfl}=3.6$ and values of S from
Fig.~\ref{Strends-fig}. Since the fraction of area occupied by the
fibrils is $1/\lambda$, $S_0$ is the local stress within the fibrils.

\begin{figure}[t]
\begin{center}
\includegraphics[width=8cm]{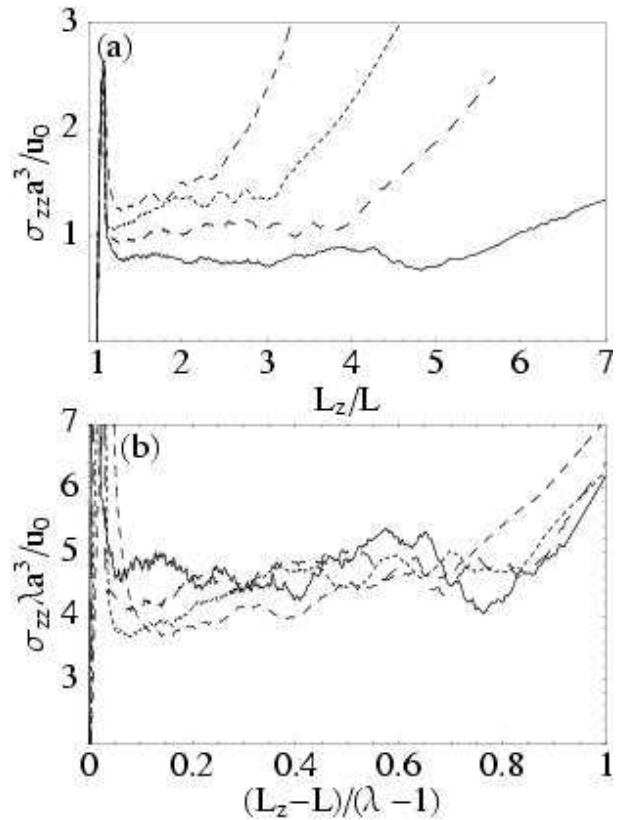}
\caption[Stress-strain curves for crazes with different values of
$l_p$]{\label{morelambda-fig}(a) Stress-strain curves for crazes with
$l_p=1.65a$, $2.2a$, $2.7a$, and $3.3a$ in order of increasing height
at $T=0.1u_0/k_B$, $r_c=1.5a$ and $b=0u_0$. The corresponding values
of $\lambda$ are given in Table \ref{Ne-tab}.
(b) Rescaling
of the same data in the form $\sigma_{zz}\lambda$ versus
$(L_z/L-1)/(\lambda-1)$. All curves collapse onto a common plateau
$S_0=S\lambda\approx 4.5u_0/k_B$ of the same length.}
\end{center}
\end{figure}

It is perhaps surprising that the value of local stress needed to draw
fibrils is independent of the entanglement length.  In order to
further test Eq.~(\ref{Slambda-eq}), it would be desirable to consider
additional values of $N_e$ and thus $\lambda$. Unfortunately, reliable
values for $N_e$ exist for only a few values of $b$. It is, however,
not necessary to know $N_e$ for the present purpose, since both $S$
and $\lambda$ can be measured directly from the craze simulation. 
Moreover the entanglement length should only depend upon the chain
statistics \cite{Fetters1994}, and glassy states with arbitrary
statistics can easily be created. We confirmed that simulations with
the same persistence length in the undeformed glass gave the same
values of $\lambda$ and $S$ independent of whether the bond-bending
potential (Eq.~(\ref{angle-eq})) was
included. Fig.~\ref{morelambda-fig}(a) compares stress strain curves
for $b=0u_0$ at four values of the initial persistence
length. Increasing $l_p$ lowers $\lambda$ and increases $S$. However,
Fig.~\ref{morelambda-fig}(b) shows that all curves can be collapsed if
$\sigma_{zz}$ is scaled by $\lambda$ and the extension by
$\lambda-1$. This confirms that $S_0$ is the stress that controls
craze growth and only depends on the van der Waals interactions. An
experimental version of this test would be difficult, since it
is hard to change $N_e$ for real polymers without changing the
chemistry as well.

It is interesting to compare values of $S_0$ to the stresses required
for shear yielding and cavitation, which are also independent of chain
statistics. Table \ref{stresscomp-fig} compares these three stresses
for two ranges of the LJ potential. Here $b=0u_0$, but the
bond-bending potential has little effect on the values. The three
stresses are clearly correlated, decreasing with increasing
temperature and decreasing $r_c$.  The local fibril stress is always
about twice the cavitation stress and ranges from 7 to 12 times
$\tau_0$.  The implication is that local stress for drawing material
into fibrils $S_0$ is related to the bulk yield stresses, but it is
difficult to determine the relative role of shear and cavity growth.
It is interesting to note that the experimental values of $S_0/\tau_0$
for PS and PMMA are about 5 \cite{Kramer1983,Quinson1997}, and it
would be useful to have values of this ratio for other polymers. A
comparison to $p_{\rm cav}$ would also be interesting, but its value
is sensitive to system size, strain rate and inhomogeneities and it is
difficult to measure experimentally.

\begin{table}[t]
\caption{\label{stresscomp-fig}Values of the shear yield stress
$\tau_0$, the yield stress for cavitation $p_{\rm cav}$ and the local
fibril stress $S_0$ as a function of temperature. Stresses are in
units of $u_0/a^3$. Uncertainties are $\pm 0.02$ in $\tau_0$, and
about 10\% in the other quantities.}
\begin{tabular}{|l|ccc|ccc|}\hline
& & $r_c=1.5 a$ & & & $r_c=2.2 a$ &\\ 
$T k_B/u_0$ & 0.3 & 0.1 & 0.01 & 0.3 & 0.1 & 0.01 \\ \hline
$\tau_0$ &  0.23 & 0.49 & 0.72 & 0.45 & 0.64 & 0.83 \\
$p_{\rm cav}$ & 1.2 & 2.7 & 3.0 & 3.0 & 4.8 & 5.0 \\
$S_0$ & 2.9 & 4.5 & 5.2 & 5.9 & 8.0 & --- \\\hline
\end{tabular}
\end{table}

Given that previous results for $N_e$
\cite{Kremer1990,Puetz2000,Faller2000} are consistent with values
inferred from the extension ratio (Eq. \ref{lambda-eq}), our results
for $\lambda$ as a function of $l_p$ allow a rapid estimation of
$N_e$.  Table \ref{Ne-tab} presents results for a wide range of $l_p$
and shows that the product $N_e l_p$ is constant within our errorbars.
Fetters {\it et al.} \cite{Fetters1994} have presented a model for the
relation between chain statistics and $N_e$ that predicts
\begin{equation}
\label{fetters-eq}
N_e\propto \frac{N^3}{\langle R^2\rangle^3}
\end{equation}
where $\langle R^2\rangle=l_pl_0N$ denotes the average end-to-end
vector of the polymer chain and the proportionality constant only
depends upon density.  This implies $N_e \propto l_p^{-3}$, while our
data is clearly consistent with a simple inverse relation $N_e \propto
l_p^{-1}$.  Equation (\ref{fetters-eq}) describes many experimental
polymers, but it is difficult to change $l_p$ without changing all the
other parameters in the equation.  The flexible model ($b=0u_0$) is
known to be quantitatively inconsistent with Equation (\ref{fetters-eq})
\cite{Puetz2000}, which has been one motivation for studies of more
rigid models.  It would be interesting to have additional values of
$l_p$ from melt simulations to test whether the inverse relation
between $N_e$ and $l_p$ found here holds more generally, and, if so,
to understand its origin.

\begin{table}[t]
\caption{\label{Ne-tab} Measured values of $\lambda$, with uncertainties,
as a function of $l_p$ and the corresponding range of values of $N_e$
and $l_p N_e$ inferred from Eq. \ref{lambda-eq}.
Runs were made at $T=0.1u_0/k_B$, $r_c=1.5a$ and $b=0u_0$.
}
\begin{tabular}{|cccc|}\hline
$l_p $ & $\lambda$ & $N_e$ & $l_p N_e$  \\ \hline
1.65 & $6.0 \pm 0.6$ & 50-76 & 83-124 \\
2.2 & $4.5 \pm 0.5 $& 37-57 & 81-127 \\
2.7 & $3.5 \pm 0.3 $& 29-41 & 78-111 \\
3.3 & $3.0 \pm 0.3 $& 25-37 & 83-122 \\
5.55 &$ 2.0 \pm 0.2$ & 19-28 & 105-155 \\\hline
\end{tabular}
\end{table}

\subsection{Width of the active zone}
At the interface between dense polymer and craze, polymer chains are
locally mobilized and brought into the new fibril structure. The
region in which this motion takes place is called the 'active
zone'. In Fig.~\ref{capmodel-fig} the height of the active zone $h$
was defined as the distance between undeformed polymer layer and the
void ceiling, and this layer was assumed to behave like a
strain-softened fluid. The main drop in density should occur over the
height of the void ceiling of order $\sim \langle D_0\rangle/2$
\cite{Kramer1990}. In this section, we compare this simple picture to
our simulations.

\begin{figure}[t]
\begin{center}
\includegraphics[width=8cm]{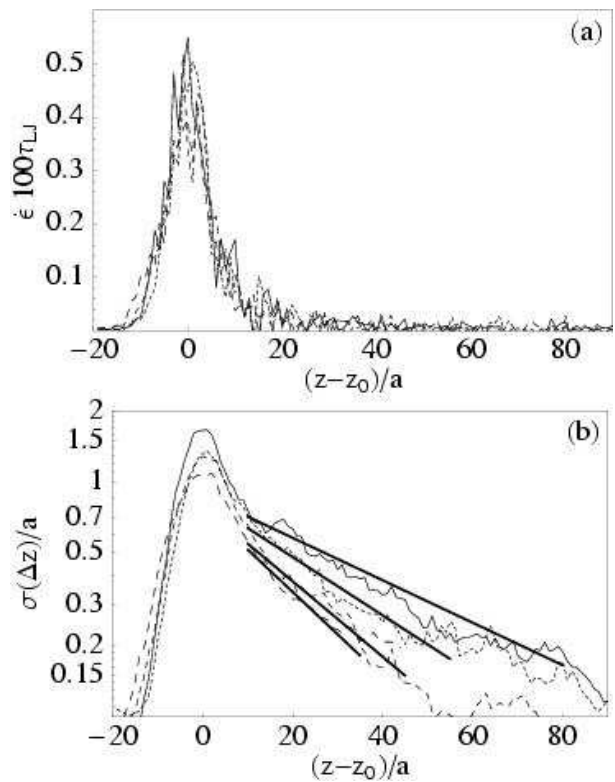}
\caption[Analysis of displacements at the active
zone]{\label{activezone-fig}(a) Strain rate $\dot{\epsilon}$ as a
function of distance from the location of the onset of density drop
$z_0$ for 4 values of $N_e$ (see text). Positive values of $z-z_0$
measure the distance into the crazed region.  The length of the dashes
increases with decreasing $N_e$. (b) Standard deviation $\sigma(\Delta
z)$ from the average displacements along z in a given layer over a
time interval of $75\,\tau_{\rm LJ}$. $\sigma(\Delta z)$ decays
exponentially into the craze and the straight lines show fits to this
decay with characteristic length scales $48a$, $35a$, $27a$, and
$24a$.}
\end{center}
\end{figure}

Fig.~\ref{densitydrop-fig} shows typical results for the density
profile near the craze boundary.  For both flexible and semiflexible
polymers, the density drops over a region of width $\sim 20 a$. The
average strain rate $\dot{\epsilon}$ must be localized in the same
region, since $\dot{\epsilon}=-\partial\ln{\rho}/\partial t$.  If the
active zone advances at velocity $v$ then $\dot{\epsilon}=\pm v
\partial\ln{\rho}/\partial z$ (the sign depends on whether the top or
bottom interface is growing). In Fig.~\ref{activezone-fig}(a) we
present $\dot{\epsilon}$ as a function of position along $z$. Averages
were taken over layers of height 1a. The curves shown correspond to
the 4 different values of $l_p$ used in the creation of the polymer
chains. As discussed above, this varies the entanglement length
$N_e$. Curves were shifted by $z_0$, which corresponds to the center
of the peak. At this point, the density is roughly halfway between
$\rho_i$ and $\rho_f$.  Curves for very different $N_e$ essentially
overlap with a width of $\sim 10a$ at half maximum and $\sim 20a$ at
the base.  The latter corresponds to the range of rapid density change
in Fig.~\ref{densitydrop-fig}.

The width of the region over which beads are mobilized was determined
from the relative diffusion as a function of height.
Fig.~\ref{activezone-fig}(b) shows the standard deviation of
displacements in the lengthening direction $\sigma(\Delta z)$. The
curves peak at the same location as the curves in
Fig.~\ref{activezone-fig}(a), but are more asymmetric. In the
direction of the dense polymer glass $\sigma(\Delta z)$ and
$\dot{epsilon}$ fall to zero over a comparable range.  By contrast,
$\sigma(\Delta z)$ shows a long exponential tail into the craze with a
characteristic decay length $l_{\rm AZ}$ that varies with entanglement
length. The fit values of $l_{\rm AZ}$ indicate that there is a
definite trend to larger values as $N_e$ increases, and $l_{\rm AZ}$
tends to be somewhat smaller than $N_e$. This result is not
surprising, because $N_e$ is the longest length scale over which
particle mobilization should occur. Standard deviations of the lateral
displacements $\Delta x$ and $\Delta y$ are smaller, but show
essentially the same decay lengths.

The above analysis indicates that while the mobility of the beads is
constrained by entanglements, the regions of localization of strain
rate and the density drop are related to the craze microstructure (see
Section \ref{structsec}) . Typical values for $\langle D\rangle$ and
$\langle D_0\rangle$ are given in Table \ref{struct-table}. From this,
$\langle D_0\rangle/2\sim 10 a$, which compares well to the width of
the strain localization peak at half maximum.  Experimentally, the
width of the active zone has been measured by a gold decoration
technique \cite{Kramer1990}. It was concluded that it lies between
$\langle D\rangle$ and $\langle D_0\rangle$, which agrees with our
results.

\section{Microstructure of Crazes}
\label{structsec}
Another fascinating aspect of crazes is their complicated
microstructure. Figs.~\ref{seq1-fig} - \ref{seq3-fig} give an
impression of the range of length scales appearing in the voided
fibril network. Clearly, the picture of cylindrical fibers and void
fingers (Fig.~\ref{capmodel-fig}) is an oversimplification. It is
nevertheless helpful to build more realistic models starting from this
simple scenario.

\subsection{Structure factor}
\begin{figure}[t]
\begin{center}
\includegraphics[width=8cm]{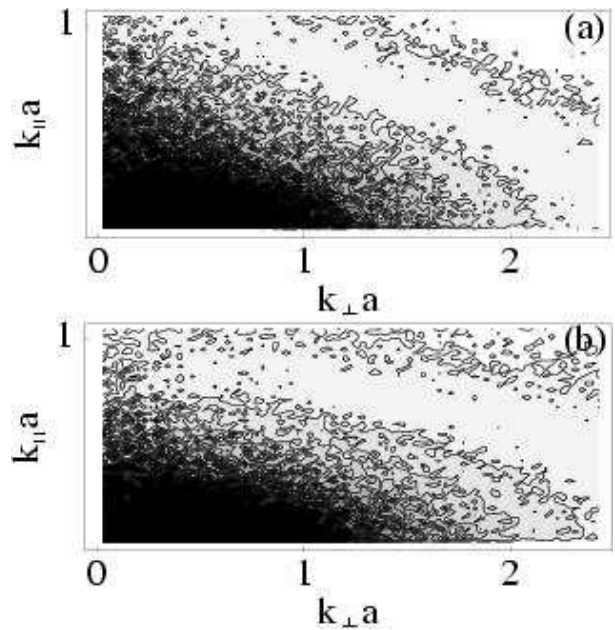}
\vspace{0.1cm}
\caption[Contour plots of the structure factor of crazes]{Contour
plots of the structure factor of crazes with (a) flexible chains,
$r_c=1.5 a$ and (b) semiflexible chains, $r_c=2.2a$. The temperature
was $T=0.1 u_0/k_B$ and the systems contained 1048576 beads. Colors
range from black (high intensity) to white (low intensity).}
\label{sofk2d-fig}
\end{center}
\end{figure}

Experimentally, the standard measurement of the craze microstructure
is done via scattering experiments. The scattering intensity in these
measurements is proportional to the structure factor
\begin{equation}
S({\bf k})=\frac{1}{N}\left\langle\sum_{j,k}e^{-i{\bf k}({\bf r}_j-{\bf
r}_k)}\right\rangle
\end{equation}
times the form factor for the monomers.  Since the craze structure has
azimuthal symmetry, one decomposes the wavevector ${\bf k}$ into
components parallel and perpendicular to the fibrils. Contour plots of
$S(k_\perp,k_\parallel)$ for two crazes are shown in
Fig~\ref{sofk2d-fig}. The microstructure was varied by changing the
cutoff length $r_c$ and chain flexibility. Both patterns are
asymmetric with the intensity decaying much faster in the direction
parallel to the fibrils than perpendicular to them.

\begin{figure}[t]
\begin{center}
\includegraphics[width=8cm]{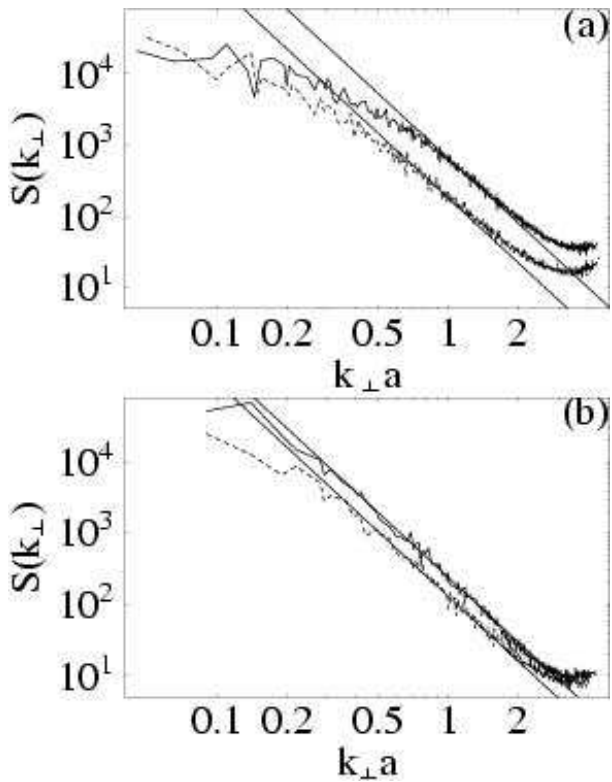}
\vspace{0.1cm}
\caption[Integrated structure factor of crazes]{Integrated structure
factor of crazes (a) at $T=0.1 u_0/k_B$, flexible chains, $r_c=1.5 a$
(solid) and semiflexible $r_c=2.2a$ (dashed) and (b) at $T=0.3
u_0/k_B$, flexible (solid) and semiflexible (dashed) chains with
$r_c=2.2 a$. The straight lines have slope -3. System sizes were
1048576 beads in (a) and 262144 beads in (b).}
\label{sofk-fig}
\end{center}
\end{figure}

Most experimental setups integrate over $k_\parallel $ using slit
collimation and measure the integrated structure factor $S(k_\perp
)=\int_\infty dk_\parallel S({\bf k})$.  In Fig.~\ref{sofk-fig}(a), we
plot $S(k_{\perp})$ as a function of the magnitude of $k_{\perp}$ for
the same systems shown in Fig.~\ref{sofk2d-fig}.  At large
wavevectors, the curves rise to a peak at $\sim 2\pi/a$ (not shown),
which corresponds to the typical separation of two beads. This length
scale is so short that it is usually not resolved in typical
experimental scattering plots shown in e.~g. \cite{Brown1987}. The
characteristic feature of $S(k_{\perp})$ is found at smaller $k_\perp$
in form of a power-law regime with exponent $-3$. The extent of this
scaling regime is bound at large wavevectors by the small scale cutoff
provided by the interparticle spacing and at small wavevectors by the
distance to the next fibril.  The power-law regime is more pronounced
for the crazes shown in Fig.~\ref{sofk-fig}(b), which were created at
higher temperatures where $\langle D_0\rangle$ is bigger.

\subsection{Interpretation of scattering data}
The traditional interpretation of craze scattering data begins with
idealizing a craze fibril as a straight cylinder of diameter $D$ and
length $l$ along $z$. The scattering intensity is then proportional to
the squared magnitude of the form factor for such a cylinder
\cite{Brown1987},
\begin{equation}
\label{cyl-eq}
F(k_{\perp})=\frac{\Delta\rho_{el}\pi D^2l^{1/2}}{2}\frac{J_1(\pi
Dk_{\perp})}{\pi Dk_{\perp}},
\end{equation}
where $\Delta\rho_{el}$ denotes the electron density and $J_1$ is the
first-order Bessel function. Due to the asymptotic behavior of
$J_1(x)=(2/\pi x)^{1/2}\cos[x-3\pi/4]+\mathcal{O}(x^{-1})$ for large
arguments, the scattering intensity of a single cylinder will exhibit
an oscillating power-law behavior, $|F(k_{\perp})|^2 \propto
k_{\perp}^{-3}$. This is also called {\em Porod} scattering.

In general the fibrils do not have a single diameter, but rather a
diameter distribution $P(D)$. One can introduce an average scattering
intensity
\begin{equation}
I_0(k_{\perp})=\langle F(k_{\perp})^2 \rangle =\int_{D_{\rm
min}}^{D_{\rm max}}dD P(D)F(k_{\perp})^2.
\end{equation}
by averaging the form factor over the diameter distribution and
neglecting correlations between fibrils.  The main effect is to smooth
out the oscillations so that a straight power-law tail results. The
average diameter $\langle D \rangle=\int DP(D)dD$ can be obtained via
a Porod analysis, in which one determines the prefactor $\alpha$ to
the power-law tail, $S(k_{\perp})=\alpha k_{\perp}^{-3}$. This can be
related to $\langle D\rangle$ \cite{Brown1981,Paredes1979} through,
\begin{equation}
\langle D\rangle=\frac{Q}{\pi^3(1-1/\lambda)\alpha},
\end{equation}
where $Q=\int dk_{\perp} 2\pi k_{\perp}S(k_{\perp})$ is a scattering
invariant. Values for $\langle D \rangle$ obtained from this formula
are collected in Table \ref{struct-table}.

The craze fibrils do not all have the same distance from each other,
but have in general varying distances that can be described by a
radial distribution function $g(r)$. This will lead to interference
effects in the scattering intensity that can be described by an
interference function $j(k_\perp)=I(k_\perp)/I_0(k_{\perp})-1$. The
interference function is related to $g(r)$ by \cite{Brown1987}
\begin{equation}
\label{gofr-eq}
g(r)=1+\frac{\pi\langle D\rangle^2\lambda}{4}\int_0^\infty2\pi k_\perp
j(k_\perp)J_0(2\pi rk_\perp)dk_\perp.
\end{equation}
As a result, the power-law tail will be modified at small
wavevectors. In particular, the first peak in $g(r)$, corresponding to
a typical fibril separation $\langle D_0\rangle$, should translate
into a maximum in $I(k_\perp)$.
\begin{figure}[t]
\begin{center}
\includegraphics[width=7cm]{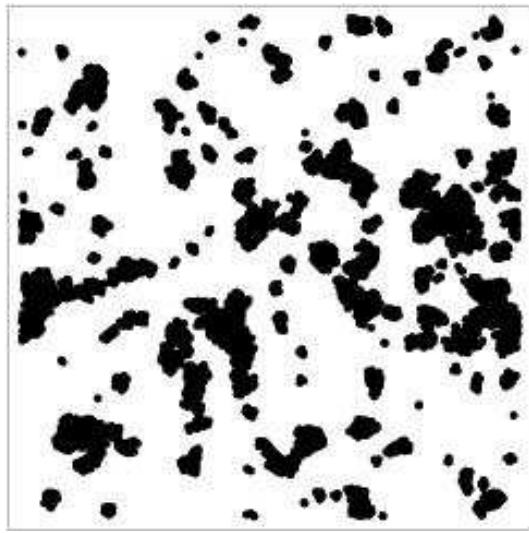}
\caption[Crosssection through a typical craze]
{\label{crosssec-fig}Typical crosssection through a craze from flexible
chains, $r_c=1.5\,a$, $T=0.1\, u_0/k_B$. The lateral dimension is
$128\,a$. Fibrils appears as clusters of varying size, and the
distributions in Fig.~\ref{fibdist-fig} are calculated form these
crosssections (see text).}
\end{center}
\end{figure}

\begin{figure}[tb]
\begin{center}
\includegraphics[width=8cm]{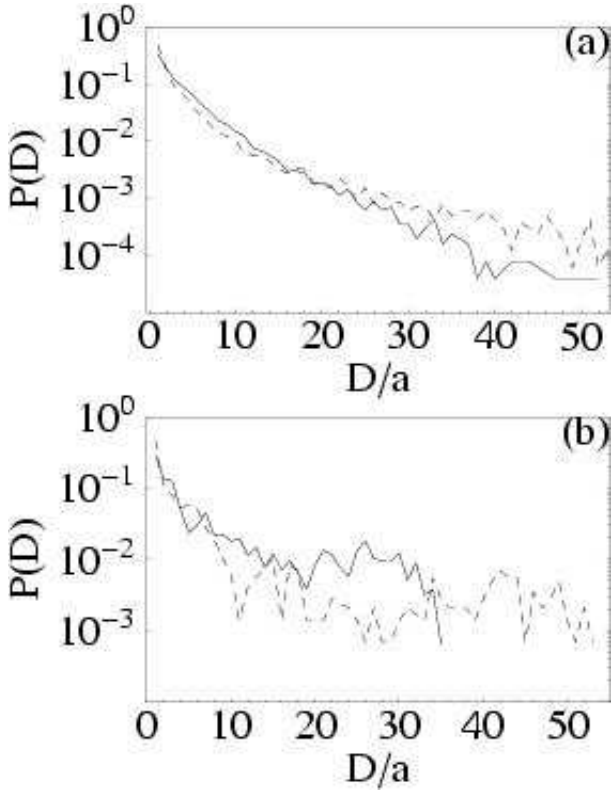}
\caption[Distribution of fibril
diameters]{\label{fibdist-fig}Distribution of fibril diameters from
analysis of connected clusters (see text) for the same crazes as in
Fig.~\ref{sofk-fig}(a) and (b). Solid lines refer to flexible chains,
dashed lines to semiflexible chains.}
\end{center}
\end{figure}

The Porod scattering law $|F(k_{\perp})|^2 \propto k_{\perp}^{-3}$ is
well confirmed at higher temperatures in Fig.~\ref{sofk-fig}(b), while
the power-law regime is shorter at lower temperatures. Visual
inspection of the craze images suggests that the straight cylinder
approximation is not so well justified in this case. Fibrils branch
more often and intersect the z-axis at varying angles. At the higher
temperature, the chains are more mobile and can align more easily, but
they are still not ideal cylinders.  We note furthermore that the
curves shown in the log-log plot of Fig.~\ref{sofk-fig} do not exhibit
a clear maximum (a maximum would be more easily identifiable in a
linear plot normally used for experimental results). This would
suggest that the ordering of fibrils is mostly random without a clear
characteristic separation. However, our statistics are limited by the
system size at these large length scales.

\subsection{Distributions of fibril diameter and spacing from real space analysis}
In a previous analysis of scattering data \cite{Brown1981} for
polystyrene and polycarbonate crazes,
Eqs.~(\ref{cyl-eq})-(\ref{gofr-eq}) were used to extract the diameter
distribution $P(D)$ and $g(r)$ by means of a detailed fitting
procedure.  The craze images shown above suggest that $P(D)$ is rather
broad, and wide distributions for $P(D)$ were found for both
materials, with a significant increase in breadth for polycarbonate
\cite{Brown1987}.  

Here, we access these distributions by direct geometrical analysis of
the bead positions.  To this end, we bin the particle positions onto a
square grid with gridsize $1 a$ normal to the widening direction and
take lateral cross-sections of height $1.5 a$. As illustrated in
Fig.~\ref{crosssec-fig}, a fibril now appears as a 2-dimensional
connected cluster, whose area $A$ is taken to be the sum of the areas
of the occupied squares. We define $D=\sqrt{4A/\pi}$.

Fig.~\ref{fibdist-fig} shows the resulting distributions of $D$. As
expected, $P(D)$ is very broad.  The distributions for flexible and
semiflexible chains are very similar at small $D$ for a fixed
temperature, but differ for larger $D$. The tail of the distributions
could be fitted to an exponential function, but our statistics are too
small for a conclusive statement.
\begin{table}[tb]
\begin{center}
\caption[Structural parameters of model
crazes]{\label{struct-table}Structural parameters of model
crazes. Size refers to the total number of beads in the
simulation. For the fibril spacing, results for $\langle D\rangle$
from both the scattering analysis and the cluster analysis (see text)
are shown. The rms variation $\sigma(D)/a$ was obtained from cluster
analysis.}
\begin{tabular}{lllccccc}
 & $T$ & $r_c$ & size & $\langle D\rangle /a$ & $\langle D\rangle /a$ & $\sigma(D)/a$ & $\langle D_0\rangle/a$ \\
 & & & & scatt. & cluster & & \\
\hline
sfl. & 0.3 & 1.5 & $128^3$ & 6.1 & 5.1 & 9.4 & 18.4 \\ 
sfl. & 0.3 & 2.2 & $128^3$ & 11.3 & 6.8 & 13.1 & 25.0  \\
fl. & 0.1 & 1.5 & $64 \times 128^2$ & 4.7 & 4.2 & 4.1 & 14.3  \\ 
sfl. & 0.1 & 1.5 & $64 \times 128^2$ & 4.8 & 4.0 & 5.5 & 12.8  \\
fl. & 0.1 & 2.2  & $64 \times 128^2$ & 8.4 & 5.5 & 7.0 & 22.3  \\ 
sfl. & 0.1 & 2.2 & $64 \times 128^2$ & 8.2 & 5.3 & 9.5 & 19.8  \\
fl. & 0.3 & 2.2 & $64^3$ & 12.6 & 7.9 & 8.7 & 30.7  \\ 
sfl. & 0.3 & 2.2 & $64^3$ & 11.1 & 6.5 & 10.4 & 23.5   \\ 
\end{tabular}
\end{center}
\end{table}
Mean values of the diameter are given in Table \ref{struct-table}
together with the standard deviations of the distributions. The large
values of the latter suggest that $\langle D\rangle$ has to be used
with care when describing the craze microstructure. Previously, Baljon
and Robbins reported similar values of $\langle D\rangle=7a$, $\sigma(D)= 11 a$
for flexible chains at $T=0.3 u_0/k_B, r_c=1.5 a$ \cite{Baljon2001}.
\begin{figure}[t]
\begin{center}
\includegraphics[width=8cm]{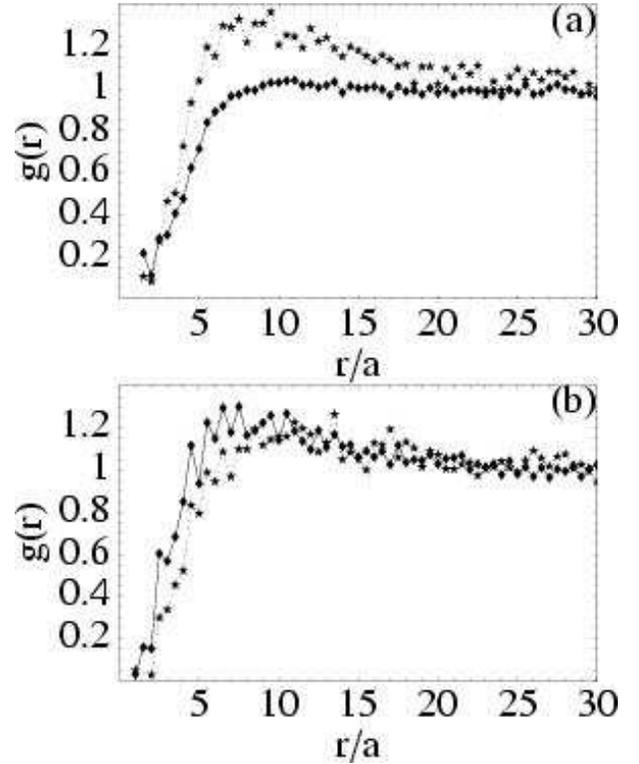}
\caption[Radial distribution function $g(r)$ of craze
fibrils]{\label{gofr-fig}Radial distribution function $g(r)$ from
analysis of connected clusters (see text).  The systems shown are (a):
$T=0.1u_0/k_B$, $r_c=1.5a$, flexible  ($\blacklozenge$) and $r_c=2.2a$, semiflexible ($\star$), (b): $T=0.3u_0/k_B$, $r_c=1.5a$ $
(\blacklozenge)$ and $r_c=2.2a$ $ (\star)$ (both semiflexible).}
\end{center}
\end{figure}

The value of $\langle D\rangle$ obtained from the Porod analysis is
always larger than the value from the cluster (real-space)
analysis. Both values rise with increasing adhesive interaction and
increasing temperature as expected. Taking $a\sim 0.8$nm, diameters
in our model crazes would correspond to a range of $\langle
D\rangle=1.4 - 6.3$nm, which is at the small end of the experimental
range. The reason is that an artificially small value of the surface
tension and a high widening velocity $v$ are used here. Both lower
$\langle D\rangle$, which allows us to use smaller system sizes. 

An estimate for the mean fibril spacing $\langle D_0\rangle$ can be
obtained by equating the area per fibril, $\pi D_0^2/4$, to the
inverse areal density $1/n$, i.e. $\langle D_0\rangle
=2\sqrt{1/n\pi}$. The areal density was obtained by counting the
number of separate fibrils per cross-section.  Values for $\langle
D_0\rangle$ are also given in Table \ref{struct-table} and translate
into a range between 10 nm and 25 nm. The higher numbers are
comparable to experiment and are obtained with $r_c=2.2 a$, which
produces more realistic surface energies.

In order to obtain the radial distribution function of the fibrils, we
continued the analysis described above and calculated the center of
mass for each 2D cluster. The positions given by this procedure were
used to calculate $g(r)$ in Fig.~\ref{gofr-fig}. In general, these
functions have very little structure. There is a size exclusion
minimum at the origin, and the curves have a weak first maximum around
$10a$. As the fibrils become thicker, the location of the maximum
shifts to larger values. Qualitatively similar curves were obtained
from experiment \cite{Brown1987}, which confirms the basically random
nature of fibril positions. The height of the maximum is too small to
be reflected in the scattering intensity.

\section{Stress Distribution and Craze Breakdown}
\label{breaksec}
In regime III of the stress-strain curve of
Fig.~\ref{stressstrain-fig} the entire volume of the simulation cell
has been converted to a craze. Elongation past the extension ratio
causes uniform straining of the craze and eventually leads to craze
failure.  Studies of this regime are directly relevant to crack
propagation in glassy polymers (Fig.~\ref{geom-fig}). The stress in
the craze region rises from $S$ at the active zone to a maximum value
$S_{\rm max}$ at the crack tip.  The elastic properties of the craze
determine the rate at which the stress rises with distance, and
$S_{\rm max}$ determines how large the craze region can become before
the crack propagates \cite{Brown1991}. These properties were recently
obtained from MD simulations and combined with continuum theory to
predict the macroscopic fracture energy \cite{Rottler2002a}. Here we
focus on the microscopic stress distribution and its relation to
$S_{\rm max}$.

\subsection{Disentanglement versus chain scission}
The craze can fail by two different mechanisms that depend on the
chain length $N$: short chains can disentangle, while very long chains
fail through chain scission \cite{Rottler2002a}. Both limiting
behaviors and the crossover between them can be addressed through our
simulations.  As can be seen in Fig.~\ref{stressstrain-fig}, short
chains of length $N=128$ form crazes that grow at the constant plateau
stress $S$, but the stress drops monotonically to zero upon straining
past $\lambda$. For longer chains, the stress $\sigma_{zz}$ rises to a
maximum value $S_{\rm max}$ that exceeds $S$.

\begin{figure}[t]
\begin{center}
\includegraphics[width=8cm]{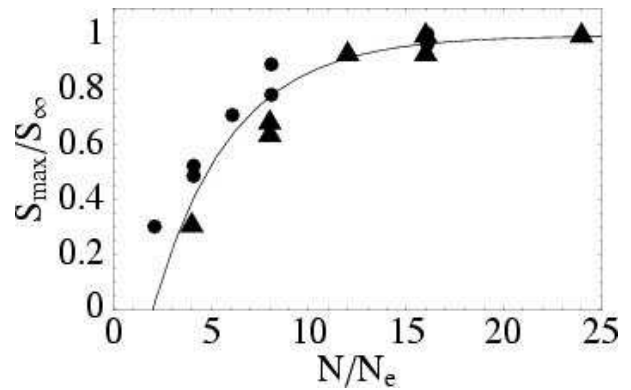}
\caption[Saturation of fibril breaking
stresses]{\label{smaxsat-fig}Saturation of fibril breaking stresses in
systems of size 262144 beads, $T=0.1 u_0/k_B$. $S_{\infty}$ denotes
the maximum saturation stress in the limit of very long chains. The
solid line is $1-\exp{(-N/4N_e +1/2)}$. Squares indicate flexible
chains ($N_e\simeq 64$) and triangles indicate semiflexible chains
($N_e\simeq 32$).}
\end{center}
\end{figure}

Values for $S_{\rm max}$ were systematically obtained as a function of
normalized chain length $N/N_e$ from curves such as those shown in
Fig.~\ref{stressstrain-fig}.  Fig.~\ref{smaxsat-fig} summarizes the
breaking stresses for the craze fibrils normalized by the breaking
stress in the limit of very long chains $S_\infty$. $S_{\rm max}$ is
zero for $N<2N_e$, since stable crazes do not form for such short
chains.  $S_{\rm max}/S_\infty$ first rises roughly linearly with
$N/N_e$, and then saturates at unity for chain lengths longer than
about $10 N_e$.  The saturation coincides with the observation of
significant amounts of chain scission.  Interestingly, the data seems
to collapse onto a single curve (solid line).

Note that the maxima of the stress-strain curves in
Fig.~\ref{stressstrain-fig} are reached at strains of $\sim 6$ and
$\sim 10$ for flexible and semiflexible chains, respectively. These
values are close to $\sqrt{3}\lambda_{\rm max}$, which implies that at
the breaking point the chains are pulled fully taut between
entanglement points. This was confirmed by direct analysis of the
craze microstructure.

\subsubsection{Chain end relaxation}
\begin{figure}[t]
\begin{center}
\includegraphics[width=8cm]{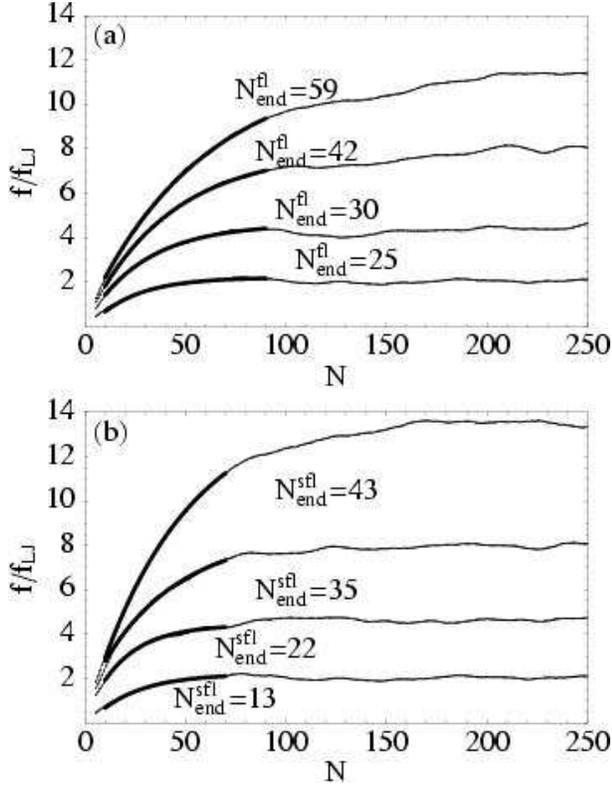}
\caption[Distribution of tension along the individual
chains]{\label{endrelax-fig}Distribution of tension along the (a)
flexible and (b) semiflexible chains (N=512). The lowest curves
corresponds to the unstressed craze and the highest curves show the
tension at the breaking point. Two intermediate stages are also shown.
The characteristic length scales $N_{\rm end}$ describe the end
relaxation and were obtained by fitting the indicated part of the
curves to an exponential relaxation.}
\end{center}
\end{figure}

In order to understand the crossover regime and the competition
between the two failure mechanisms, it is useful to study the
distribution of tension along a given chain. Fig.~\ref{endrelax-fig}
shows the tension as a function of distance from the chain end at
several stages of craze breaking.  Since the chain ends are identical,
symmetry was used to improve statistics. In the unstrained craze
(lowest curves), both flexible and semiflexible systems exhibit a
constant stress in the center of the chain, but a relaxation toward
the free ends. The characteristic length scales for this relaxation
were extracted by fitting an exponential decay to the transition
region. The values of the decay lengths $N_{\rm end}^{\rm fl}=21$ and
$N_{\rm end}^{\rm sfl}=13$ are comparable to the characteristic length
of $N_e/3$ for stretched segments, but are not universal. Stronger
adhesive interactions were found to increase $N_{\rm end}^{\rm
(s)fl}$.
Upon straining the craze, the tension in the center of the chains and
the values of $N_{\rm end}^{\rm (s)fl}$ rise. At the breaking point (last
curves), the end relaxation extends over a length scale comparable to
the entanglement length.

These results help to formulate a simple argument for the universal
curve plotted in Fig.~\ref{smaxsat-fig}. The {\em average} distance of
an entanglement point from the chain end is $N/4$. We assume that the
probability of disentanglement decreases exponentially with distance
from the chain end, as suggested by the tension relaxation curves. The
characteristic length scale $N_{\rm end}$ at the breaking point in
these curves was on the order of $N_e$. Because of the above, we
expect this length scale to be the characteristic decay length for the
probability of disentanglement, and thus postulate a
disentanglement probability of the form $\exp[-(N-2Ne)/4N_e]$. Here,
$N$ was reduced by $2N_e$, since for this chain length the
disentanglement probability is one and the chain is free on either
side. The maximum stress can now be written as the limiting value of
$S_{\infty}$ times the probability for non-disentanglement, which
gives
\begin{equation}
S_{\rm max}/S_{\infty}=1-\exp[-(N-2Ne)/4N_e].
\end{equation}
Fig.~\ref{smaxsat-fig} shows that this curve agrees well with the data.

\subsection{Global tension distribution}
\begin{figure}[t]
\begin{center}
\includegraphics[width=8cm]{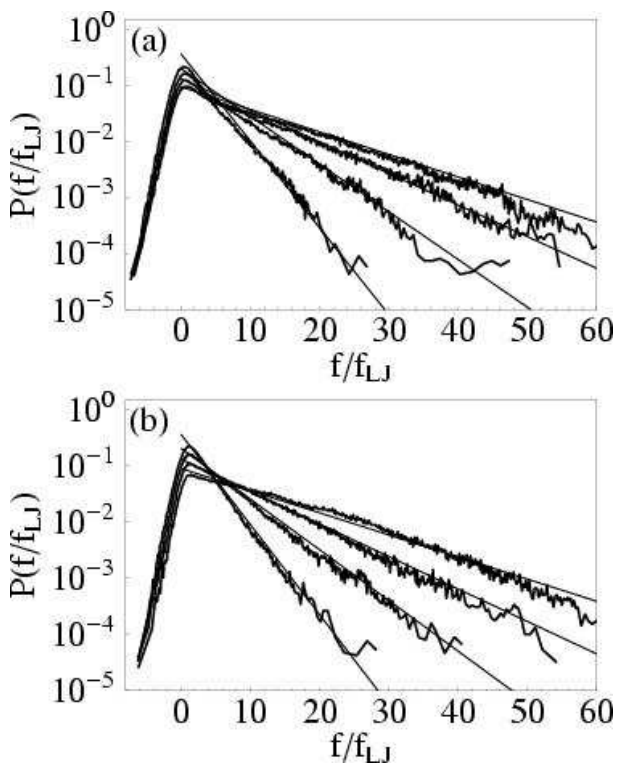}
\caption[Distribution of tension in unstressed and stressed
crazes]{\label{globalstress-fig}Distribution of tension in crazes of
size 262144 beads with N=512 for (a) flexible and (b) semiflexible
chains. Strain states correspond to the ones shown in
Fig.~\ref{endrelax-fig}. The values for $\langle f \rangle$ are
$2.9, 4.9, 7.6,$ and $11.8f_{\rm LJ}$, respectively. $f_{\rm
LJ}=2.4u_0/a^3$ is the breaking force of the LJ interaction. The
straight lines correspond to $\exp[-f/\langle f\rangle]/\langle f
\rangle$.}
\end{center}
\end{figure}

The parameter governing chain scission and thus the value of
$S_{\infty}$ is the distribution of tension in the polymer craze. In a
previous paper \cite{Rottler2002b}, we reported that this distribution
is characterized by an exponential tail at large tensile forces, in
analogy to jammed systems such as granular materials
\cite{Liu2001}. This distribution is shown in
Fig.~\ref{globalstress-fig} for flexible and semiflexible chains of
length $N=512$ and several strain states. The tensile (positive) part
of the distribution is well fitted by $1/\langle
f\rangle\exp[-f/\langle f\rangle]$, where only the positive tensions
are included in the average $\langle f\rangle$. Note that $\langle
f\rangle$ is the same for flexible and semiflexible chains at the
plateau. The distribution with the steepest slope (smallest $\langle
f\rangle$) corresponds to the fully developed craze. Additional curves
with higher $\langle f \rangle$ correspond to stressed crazes at the
same strain with respect to the unstrained craze. Note that the
semiflexible and flexible crazes have the same values of $\langle f
\rangle$ at each strain. This is related to the fact that the stress
$S_0$ in the fibrils is independent of $N_e$ (see Section
\ref{S0-sec}). The last curve shows the tension distribution at the
breaking point where $\sigma_{\rm zz}$ is largest (see also
Fig.~\ref{stressstrain-fig}). The effect of straining the craze only
changes $\langle f\rangle$, and all curves could be collapsed after
normalizing by $\langle f\rangle$.

We note that in our simulations, no scission or disentanglement occurs
during craze growth proper. The fraction of bonds that break at a
given average tension is $P_{br}(\langle
f\rangle)=\int_{f_c}^{\infty}\exp[-f/\langle f\rangle]d
f=\exp[-f_c/\langle f\rangle]$. In our simulations $f_c=100 f_{\rm
LJ}$ as described in Section \ref{modelsec}. The onset of scission
can be estimated using a simple scaling argument:
$N_{bonds}P_{br}(\langle f\rangle)\sim 1$. From this we estimate an
average value of the tension at breaking

\begin{equation}
\label{break-eq}
\langle f \rangle=f_c/\ln{N_{bonds}}. 
\end{equation}
For $N_{bonds}\simeq 262144$, this implies that scission will occur
when $\langle f\rangle \simeq 8.0 f_{\rm LJ}$, which was confirmed by
direct inspection of the chains at the corresponding strains. Such
high tensions only occur when the craze is strained past the extension
ratio. The largest value of $\langle f\rangle$ observed with the
present model during craze growth was $4.5 f_{\rm LJ}$ and occurred at
very low temperature $T=0.01 u_0/k_B$ and $r_c=2.2 a$.

The degree of chain scission in experimental crazes is still a matter
of debate, but it appears likely that at least some chains do
experience scission. The absence of scission in the present study is
most likely due to the low monomeric friction coefficient of the
bead-spring model. As the above argument showed, a relatively modest
increase in the average tension will quickly lead to appreciable
scission.  More realistic polymer models should be able to capture
this effect. An increase in system size will likewise raise the number
of broken bonds. For a typical value of $\langle f \rangle=3f_{\rm
LJ}$, one bond would break for every $10^{15}$ bonds. Note that the
exponential tension distribution leads to a logarithmic size
dependence Eq.~(\ref{break-eq}) and allows for sequential bond
breaking. The fibrils are thus much weaker than implied by the common
simple assumption that all bonds carry the same tension and break when
$\langle f \rangle=f_c$.

\section{Summary and Conclusions}
\label{concsec}
This paper presented molecular dynamics simulations of craze
nucleation, widening and breakdown.  Initial failure of the LJ polymer
glass occurred through shear in biaxial loading. Only when all three
principal stresses were tensile did cavitation and craze formation
occur. However, once past the nucleation phase, plane stress
conditions are sufficient for continuing craze growth.  Cavitational
failure could be fitted to a cavitation criterion of the form
$\tau_{dev}^{c}=\tau_0^c+\alpha^c p$.

Craze widening proceeds in the simulations by a clearly identifiable
fibril drawing process. This interpretation is also well-supported by
experiments. The resulting craze microstructure is compellingly
similar to TEM images of experimental crazes, and the length scales
quantified by $\langle D\rangle$ and $\langle D_0\rangle$ are within
experimental limits. The simulations clearly establish the connection
between extension ratio and entanglement length. In the glass,
disentanglement is prevented and the entanglements act like chemical
crosslinks. A microscopic analysis of the length of stretched chain
segments has shown that unlike the case of a regular mesh, only a few
segments are fully expanded to the entanglement length, and the
average extension is only $N_e/3$. The factor 1/3 arises from
averaging over all angles that a given segment can form with the
stretching direction.

Another salient finding of this study is the exponential distribution
of tension in the craze. The presence of large stress fluctuations
makes chain scission much more likely than e.~g.~a Gaussian
distribution or uniform loading. Since force distributions of this
kind are also often seen in conventional ``jammed'' systems such as
foams, colloids and granular media, we have suggested
\cite{Rottler2002b} that a craze can be viewed as a system that jams
under tension.

The highly nonequilibrium nature of the force distribution, and the
strong concentration of stress in the covalent backbone bonds, formed
the basis for our critique of the conventional capillary model of
craze widening. The polymer glass is not a viscous fluid in the active
zone, and the hydrodynamic description does not apply. The picture
suggested by our simulations is that crazing is a form of localized
shear deformation, but with a much greater mobilization of material
than in standard shear yielding.  The very similar rate and
temperature dependence of $S$ is another indication of the close
relation between the processes. Based on trends observed in the
simulations, we have suggested that the local stress in the fibrils
$S_0=S\lambda$ is independent of the entanglement length. $S_0$ varies
with temperature and strength of adhesive interaction in a manner
very similar to the yield stresses for shear and
cavitation. Establishing a precise connection between these
characteristic stresses should be a most interesting direction for
future work.

A detailed analysis of the microstructure of crazes was also
presented.  The calculated structure factor is similar to measured
scattering intensities.  As in these experiments, a Porod analysis was
used to extract a measure of the mean fibril diameter $\langle D
\rangle$ from the structure factor.  While the extension ratio depends
only on $N_e$, the mean fibril diameter depends on many factors.  The
value of $\langle D \rangle$ increases with increasing $T$ and with
increasing strength of the van der Waals interactions.  Chain
stiffness has less effect, although $\langle D \rangle$ is larger for
flexible chains than semiflexible chains at high temperatures.

The distribution of fibril diameters was determined from the real
space structure of the crazes.  The average fibril diameter from this
method was always smaller than that determined from the structure
factor.  The distribution was also very wide with a variance that
exceeded the mean and a tail extending to many times the mean.  The
radial distribution function for the fibrils shows almost no
correlation, merely an exclusion minimum near the origin.  Fibrils
merge and split with each other directly, rather than being joined by
smaller cross-tie fibrils.

The simulations described here capture the generic features of
experiments on many different polymers and provide previously
inaccessible information about the dynamics and microstructure.
However, they are unable to address quantitative behavior of specific
polymers.  Future studies with chemically realistic potentials will be
of great value, but require orders of magnitude more computational
effort.

\section{Acknowledgements}
We are indebted to E.~J.~Kramer and H.~R.~Brown for very insightful
discussions of this work. Financial support from the Semiconductor
Research Corporation (SRC) and NSF grant No. DMR0083286 is
gratefully acknowledged. The simulations were performed with {\em
LAMMPS 2001} \cite{Lammps2001}, a molecular dynamics package
developed by Sandia National Laboratories.

\end{document}